\documentclass[fleqn,usenatbib]{mnras}

\usepackage{newtxtext,newtxmath}
\usepackage[T1]{fontenc}

\DeclareRobustCommand{\VAN}[3]{#2}
\let\VANthebibliography\thebibliography
\def\thebibliography{\DeclareRobustCommand{\VAN}[3]{##3}\VANthebibliography}

\usepackage{graphicx}	% Including figure files
\usepackage{amsmath}	% Advanced maths commands
\usepackage[dvipsnames]{xcolor} 

\title[Final collapse of PPSNe]
{The final core collapse of pulsational pair instability supernovae}

\author[J.~Powell, B.~M\"uller, \& A.~Heger]{
Jade Powell,$^{1,3}$\thanks{E-mail: dr.jade.powell@gmail.com}
Bernhard M\"uller,$^{2,3}$
and Alexander Heger$^{2,3,4}$
\\
$^{1}$Centre for Astrophysics and Supercomputing, Swinburne University of Technology, Hawthorn, VIC 3122, Australia.\\
$^{2}$Monash Centre for Astrophysics, School of Physics and Astronomy, Monash University, VIC 3800, Australia. \\
$^{3}$ARC Centre of Excellence for Gravitational Wave Discovery (OzGrav), Melbourne, Australia. \\
$^{4}$ARC Centre of Excellence for Astrophysics in Three Dimensions (ASTRO-3D), Australia.
}

\pubyear{2021}

\begin{document}
\label{firstpage}
\pagerange{\pageref{firstpage}--\pageref{lastpage}}
\maketitle

\begin{abstract}
We present 3D core-collapse supernova simulations of massive Pop-III progenitor stars at the transition to the pulsational pair instability regime. 
We simulate two progenitor models with initial masses of $85\,\mathrm{M}_{\odot}$ and $100\,\mathrm{M}_\odot$ with the LS220, SFHo, and SFHx equations of state.  The $85\,\mathrm{M}_{\odot}$ progenitor experiences a pair instability pulse coincident with core collapse, whereas the $100\,\mathrm{M}_{\odot}$ progenitor has already gone through a sequence of four pulses $1\mathord,500$ years before collapse in which it ejected its H and He envelope.  The $85\,\mathrm{M}_{\odot}$ models experience shock revival and then delayed collapse to a black hole (BH) due to ongoing accretion within hundreds of milliseconds.  The diagnostic energy of the incipient explosion reaches up to $2.7\times10^{51}\,\mathrm{erg}$ in the SFHx model.
Due to the high binding energy of the metal core, BH collapse by fallback is eventually unavoidable, but partial mass ejection may be possible.
The $100\,\mathrm{M}_\odot$ models have not achieved shock revival or undergone BH collapse by the end of the simulation.  All models exhibit relatively strong gravitational-wave emission both in the high-frequency g-mode emission band and at low frequencies.  The SFHx and SFHo models show clear emission from the standing accretion shock instability.  
For our models, we estimate maximum detection distances of up to $\mathord{\sim}46\,\mathrm{kpc}$ with LIGO and $\mathord{\sim} 850\,\mathrm{kpc}$ with Cosmic Explorer.
\end{abstract}


%%%%%%%%%%%%%%%%%%%%%%%%%%%%%%%%%%%%%%%%%%%%%%%%%%
%%%%%%%%%%%%%%%%% BODY OF PAPER %%%%%%%%%%%%%%%%%%
\section{Introduction}
\label{sec:intro}

Core-collapse supernovae (CCSNe) occur when the iron cores of stars above $\mathord{\sim}8\,\mathrm{M}_\odot$ reach their effective Chandrasekhar mass and collapse until they reach nuclear density. As the core rebounds elastically, a shock wave is launched outwards which quickly loses energy and stalls.
For a successful explosion, the shock must be revived.  
According to the current paradigm, shock revival
is achieved by neutrino heating in most CCSNe, but in rare cases of unusually energetic ``hypernovae'' some form of magnetohydrodynamic mechanism may play a key role as well \citep[for a review, see][]{2012ARNPS..62..407J}.

Due to the complicated nature of CCSNe, simulations are essential for understanding their explosion dynamics, observable multi-messenger emission, and remnant properties.  Multi-dimensional simulations of neutrino-driven CCSNe have advanced rapidly in recent years and are starting to reveal the systematics of explosion and remnant properties, however, the full parameter space of self-consistent 3D explosions has not yet been fully explored (see \citealt{2020LRCA....6....3M,2020arXiv200914157B} for recent reviews). 

CCSNe are the birth places of neutron stars and stellar mass black holes (BHs), which are the primary sources for gravitational-wave (GW) detectors such as Advanced LIGO \citep{aLIGO}, Advanced Virgo \citep{AdVirgo} and KAGRA \citep{Somiya_2012}.  The birth masses, spins and kicks of these compact objects cannot be understood without CCSN simulations.  

Recent GW detections of high-mass BHs have drawn particular attention to the upper end of the supernova progenitor-mass distribution and the transition to different explosion regimes.
Instead of proceeding with advanced nuclear burning stages up to the formation of an iron core,
stars with helium cores in the range $\sim 50-150\,\mathrm{M}_\odot$ are believed to become unstable to electron-positron pair production \citep{1964ApJS....9..201F,barkat_67}, which may completely unbind the star resulting in a pair-instability supernova explosion, leaving no remnant behind \citep{2002ApJ...567..532H, heger_03,2014ApJ...792...44C, 2017MNRAS.464.2854K}.  Pulsational pair-instability supernovae are stars with helium core masses in the range $\mathord{\sim} 30\texttt{-}50\,\mathrm{M}_\odot$
\citep{2002ApJ...567..532H,heger_03}.  They also experience pair instability that results in pulsations that eject material, but as the energy of the pulsations is lower, the star is not completely disrupted \citep{heger_03,2017ApJ...836..244W}.  These stars are then expected to undergo a regular core collapse that may result in a supernova or a gamma-ray burst (GRB) \citep{2007Natur.450..390W}.  As a result, stars in this mass range should form BHs with masses in the range of $30\texttt{-}45\,\mathrm{M}_\odot$; above that one expects a ``mass gap'' from the complete disruption of the star by a successful pair instability supernova \citep{2016A&A...594A..97B,2019ApJ...882..121S}.
However, optical observations of binary systems \citep{2019Natur.575..618L} as well as GW signals from binary BH mergers by LIGO and Virgo \citep{2020arXiv200901075T} have recently discovered BHs with suggested or most likely masses in this pair-instability mass gap.  Simulations of the explosions of very massive BH-forming models are important to provide further insights into the unexpected masses found by GW observations. 

CCSNe are also of interest for GW astronomy as targets in their own right.
As the sensitivity of GW detectors increases, they will begin to detect not only binary mergers but also other lower-amplitude sources of GWs such as CCSNe.  Accurate knowledge of the GW emission from CCSNe will be essential for detection and parameter estimation.  The GW signal from rotational core bounce has already been well covered in the literature \citep[e.g.,][]{2008PhRvD..78f4056D,2014PhRvD..90d4001A,2015MNRAS.450..414F, 2017PhRvD..95f3019R}.  In the non-rotating case, 
the GW emission from the post-bounce phase has been studied using self-consistent 3D simulations by many groups \citep{2017MNRAS.468.2032A,andresen_19,andresen_20, kuroda_16,2017ApJ...851...62K,Kuroda_2018, 2019ApJ...876L...9R, 2019MNRAS.487.1178P, 2020MNRAS.494.4665P,2020PhRvD.102b3027M,2020arXiv201002453P}.  The structure of the GW emission has shown common features in different simulations from recent years. The dominant emission feature in the GW emission is due to the quadrupolar surface f/g-mode
\footnote{The mode that sets the dominant emission frequency can change character from a g-mode to an f-mode \citep{2018ApJ...861...10M,sotani_20}.
For the sake of simplicity, we often refer to its frequency simply as the g-mode frequency even though its precise character at a given time is not known.} of the proto-neutron star (PNS), which produces GW frequencies rising in time from a few hundred Hz up to a few kHz \citep{2012A&A...537A..63M, 2017PhRvD..96f3005S, 2018ApJ...861...10M, 2018MNRAS.474.5272T, Kuroda_2018,2019PhRvL.123e1102T}.  In addition, some models \citep{kuroda_16,2017ApJ...851...62K,2017MNRAS.468.2032A,2020MNRAS.494.4665P,2020PhRvD.102b3027M} exhibit low-frequency GW emission due to the standing accretion shock instability (SASI; \citealp{0004-637X-584-2-971, 2006ApJ...642..401B, 2007ApJ...654.1006F}).
In rapidly rotating models, very strong GW emission can also occur during the post-bounce phase due to a corotation instability \citep{takiwaki_18}.
The emerging understanding of the GW emission features has led to the formulation of universal relations for the GW emission \citep{2019PhRvL.123e1102T} and paved the way for phenomenological modelling for CCSN signals \citep{2018PhRvD..98l2002A}.
Further work is still needed, however, to extend these models to fully explore CCSN GW signals from across the progenitor parameter space. 
The majority of 3D simulations that include GW emission are for progenitor stars below $30\,\mathrm{M}_{\odot}$.  In this paper, we perform simulations of high-mass Population~III (Pop-III) stars in the pulsational pair instability regime to expand the parameter space coverage of 3D simulations and to provide further insights into the massive and very massive star remnant BH population. 

A small number of studies have already focused on failed or partially successful CCSNe with BH formation in the regime of high-mass progenitors, but they have not extensively investigated their GW emission.
\citet{Kuroda_2018} simulated the collapse to a BH of a $70\,\mathrm{M}_{\odot}$ progenitor, and found very large GW amplitudes as convection dominated over SASI in their model.
Strong SASI was also found in the simulations of a $70\,\mathrm{M}_{\odot}$ progenitor by \citet{2020MNRAS.493L.138S, 2021MNRAS.tmp..261S}.
No such phenomenon was reported in the
3D simulations of BH formation in a $40\,\mathrm{M}_{\odot}$ progenitor
by \citet{2018ApJ...852L..19C,chan_20} despite powerful SASI activity with a clear imprint
on the neutrino signal \citep{mueller_19b}, but no further analysis of the GW emission has been
carried out for these models.  The GW emission prior to BH collapse also remains of modest
amplitude in the recent 3D models of BH collapse for a $40\,\mathrm{M}_\odot$ progenitor by \citet{2020arXiv201002453P},
though an earlier 2D study \citep{2018ApJ...857...13P} did show enhanced GW emission shortly before collapse in some cases.
The $40\,\mathrm{M}_\odot$ models of \citet{2020arXiv201002453P}
with different rotation rates were, however, noteworthy for predicting very high GW frequencies of up to $\mathord{\sim}3,000$\,Hz before BH formation.  It is important to determine if this is a robust prediction because of the  strongly frequency-dependent sensitivity of GW detectors.
The recent work on BH formation in $40\,\mathrm{M}_\odot$ and $75\,\mathrm{M}_\odot$ progenitors
by \citet{2020PhRvD.101l3013W} did not discuss GW emission, but pointed out an interesting feature in BH-forming models that could lead to very strong GW emission. Due to the extreme recession of the shock, the $l=2$ quadrupole mode of the SASI becomes unstable and dominates the $l=1$ mode during some phases of the evolution. A strong $l=1$ mode already gives a relatively strong GW signal because of a finite admixture of $l=2$ density perturbations that are seen in GWs.  If the dominant SASI mode has $l=2$ to begin with, the signal could be much stronger.  

Beyond the high-mass end of the ``mass gap'' \citet{2001ApJ...550..372F} simulated the collapse of a  rapidly-rotating $300\,\mathrm{M}_\odot$ Pop-III model.  In that simulation a $50\,\mathrm{M}_\odot$ rapidly-rotating core formed during collapse that was held up by trapped neutrinos and susceptible to secular triaxial instabilities that could grow on a time-scale shorter than the collapse time, potentially also being a powerful GW source with $h_+\sim10^{-21}$ at $1\,\mathrm{Gpc}$.

If there is strongly enhanced GW emission prior to the final collapse, BH-forming massive stars may be observable in GWs at larger distances than normal CCSNe.  Such a strong GW signal could provide clues about the neutron star mass and radius before the final collapse, e.g., through the maximum g-mode frequency.  A GW detection from such an event would provide valuable complementary information about BH formation to optical surveys for disappearing massive stars  \citep{2015PASA...32...16S, 2015MNRAS.450.3289G, 2017MNRAS.468.4968A, 2017MNRAS.469.1445A}. In this context, it is intriguing that recent simulations \citep{2018ApJ...852L..19C,Kuroda_2018, 2018ApJ...855L...3O,2020arXiv201002453P} suggested that shock expansion could occur in massive stars just before BH formation; and this ``hiccup'' may even give rise to an observable transient \citep{moriya_19}.  Further simulations of BH forming models are needed to determine when to expect stars to collapse quietly and when there may be an early shock revival before collapse followed by extensive fallback.

In this paper, we aim to further clarify the fate and GW signatures of the most massive CCSN progenitors.  We perform 3D simulations of BH forming stellar collapse with three different equations of state (EoS) that result in different maximum neutron star masses and BH formation times. In addition to the LS220 EoS
\citep{Lattimer:1991nc} used in our previous simulations \citep{2019MNRAS.487.1178P, 2020MNRAS.494.4665P}, we will use two EoS (SFHx and SFHo, \citealp{2013ApJ...774...17S}) with a higher maximum neutron star mass and smaller neutron star radii.  The progenitor models are $85\,\mathrm{M}_\odot$ and $100\,\mathrm{M}_\odot$ Pop-III stars. These masses are higher than any of the other recent self-consistent 3D simulations of BH forming models and probe the lower end of the pulsational pair instability regime.
 
Using these models, we investigate the possibility of shock expansion before BH formation for a wider range of progenitors and EoS and, where applicable, examine the effects of the different EoS on the explosion dynamics.
We then analyse the detectability of the GW signals.  We determine the maximum detection distances for our models using simulated design sensitivity Gaussian noise for the LIGO, Einstein Telescope \citep{0264-9381-27-19-194002}, and Cosmic Explorer \citep{2017CQGra..34d4001A} detectors, and we also  discuss the detectability of features in the time-frequency structure of the signal in noisy spectrograms.

The outline of our paper is as follows: In Section~\ref{sec:models}, we present the two progenitor models.  In Section~\ref{sec:simulation}, we provide details on the numerical methods and the setup of our simulations.  In Section~\ref{sec:explosion}, we analyse the dynamics of our models.  We describe the features of the GW emission in Section~\ref{sec:gravwaves}, and present a discussion and conclusions in Section~\ref{sec:conclusion}.

%%%%%%%%%%%%%%%%%%%%%%%%%%%%%%%%%%%%%%%%%%%%%%%%%%
%%%%%%%%%%%%%%%%%%%%%%%%%%%%%%%%%%%%%%%%%%%%%%%%%%

\begin{figure}
\includegraphics[width=\columnwidth]{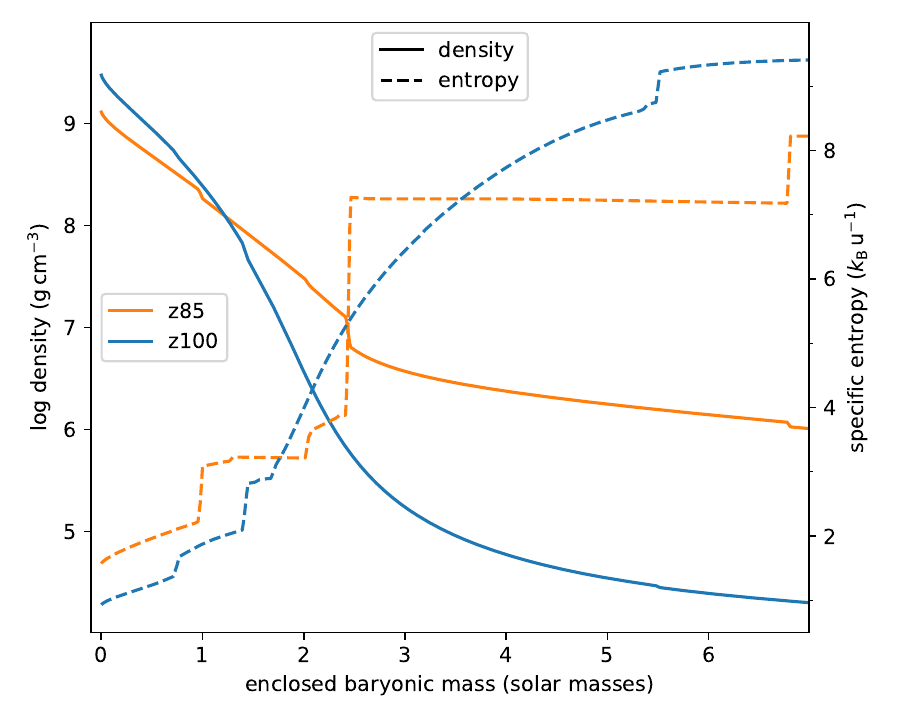}
\caption{Profiles of the density (solid lines) and specific entropy (dashed lines) as a function of mass coordinate, $m$, for the progenitor models \texttt{z85} (orange) and \texttt{z100} (blue) at the onset of core collapse, defined as the first model in which the infall velocity exceeds $900\,\mathrm{km}\,\mathrm{s}^{-1}$.  Model \texttt{z85} has a large jump in entropy and density between the silicon core and the convective O shell at a mass coordinate of $m=2.4\,\mathrm{M}_\odot$.  In model \texttt{z100}, traces of remaining oxygen burn between $m=1.9\,\mathrm{M}_\odot$ and $m=5.5\,\mathrm{M}_\odot$ leading to a smooth shallow entropy and density gradient instead.  Unlike model \texttt{z85}, the flat profile above $m=5.5\,\mathrm{M}_\odot$ is not convective and not powered by nuclear burning.  The unusual structure is a result of the final pair-instability pulse (Figures~\ref{fig:z100_khd} and \ref{fig:z100_pulse}). 
}
\label{fig:profiles}
\end{figure}

\begin{figure}
\includegraphics[width=\columnwidth]{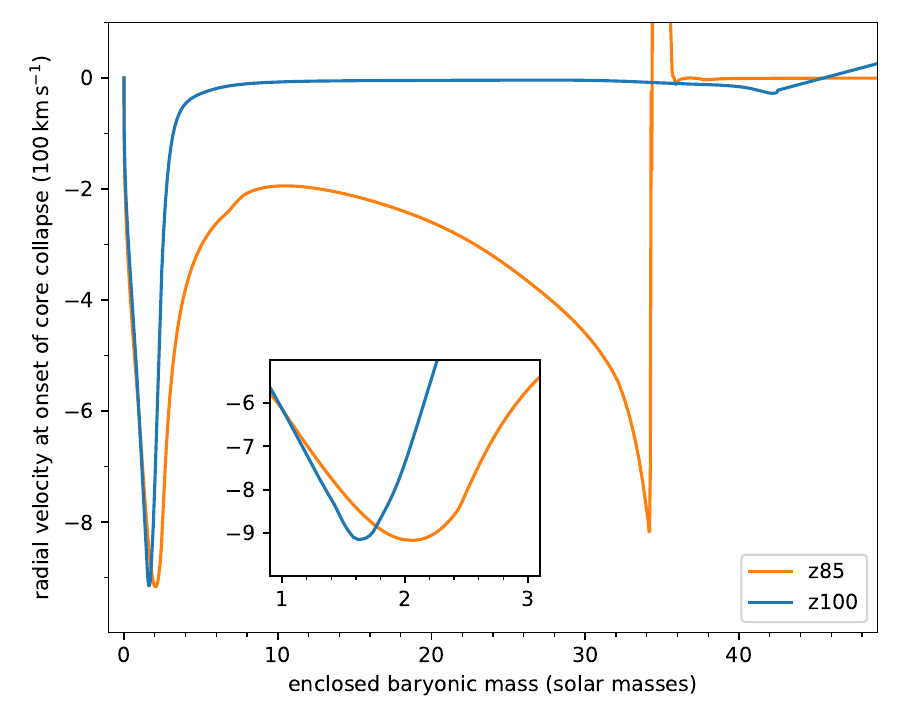}
\caption{Radial velocity profile of models \texttt{z85} (\textsl{orange}) and \texttt{z100} (\textsl{blue}) as a function of mass coordinate $m$ at onset of core collapse, defined as the first model in which the infall velocity exceeds $900\,\mathrm{km}\,\mathrm{s}^{-1}$.  The insert magnifies the locations of the peak infall velocities, which are at mass coordinates of $2.1\,\mathrm{M}_\odot$ and  $1.6\,\mathrm{M}_\odot$, respectively.  In model \texttt{z85}, the entire CO core of $\mathord\sim34\,\mathrm{M}_\odot$ is collapsing due to pair instability occurring at iron core collapse, whereas in model \texttt{z100} a much slower contracting core is seen out to $m\sim42.5\,\mathrm{M}_\odot$.  In model \texttt{z100}, the entropy in the core is much larger in the aftermath of the preceding pair-instability pulses and a much smaller ($1.6\,\mathrm{M}_\odot$) homologously-collapsing iron core results than in model \texttt{z85} ($2.1\,\mathrm{M}_\odot$). 
}
\label{fig:collapse}
\end{figure}

\section{Progenitor Models}
\label{sec:models}
\begin{figure*}
\centering
\includegraphics[width=\columnwidth]{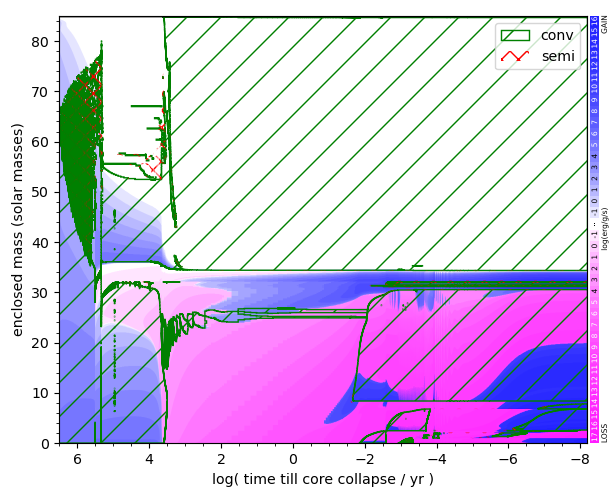}
\hfill
\includegraphics[width=\columnwidth]{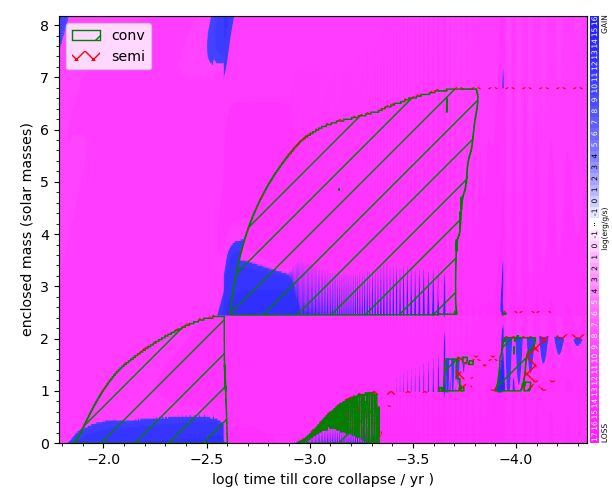}
\caption{Kippenhahn diagrams for model \texttt{z85}.
\newline
\textsl{Left panel}: The entire evolution from zero-age main sequence (ZAMS) to core collapse on a logarithmic time scale where we assume ``core collapse'' or core bounce, is reached 1/4 second after the last model shown, which is a reasonable estimate.  The $y$-axis shows the enclosed mass (mass coordinate).  \textsl{Green hatching} indicates convective regions, also outlined by a \textsl{green line}; \textsl{red cross hatching} indicates semi-convective regions.  The regions appearing solid green on the left side during core hydrogen burning around a mass coordinate of $60\,\mathrm{M}_\odot$ is effectively semi-convective, but threaded though with many small convective zones that each contribute their green outlines to the plot.  \textsl{Blue shading} indicates net specific nuclear energy generation rate, each level of increased shading intensity indicating an increase in energy generation rate by one order of magnitude, with the faintest level being $0.1\,\mathrm{erg}\,\mathrm{g}^{-1}\,\mathrm{s}^{-1}$.  \textsl{Purple shading} indicates net specific nuclear energy loss, using the same scheme as for energy generation.  For both, we plot the net value of energy generation and neutrino losses, as it is this that affects the evolution and structure of the star. Core hydrogen burning (main sequence, MS) is from the start until about $5.5$ on the $x$-axis  ($300\mathord{,}000\,\mathrm{yr}$ prior to collapse).  Below we adopt the short form ``$x=5.5$.''  Core helium burning is until about $x=3.6$, the neutrino-powered CO core contraction phase is only $\mathord\sim4\mathord,000\,\mathrm{yr}$.  There is a major core-envelope mixing event at around $x=5.4$ that leads to the establishment of a powerful convective hydrogen-burning shell between $36\,\mathrm{M}_\odot$ and $53\,\mathrm{M}_\odot$ that is enriched in CNO material from core helium burning, increasing entropy and preventing further core-envelope mixing.  Core carbon burning starts in a radiative manner less than one year before collapse, neon burning is not prominent due to low abundances of carbon and neon made in this star.  Core oxygen burning starts at around $x=-1.9$, $4$ days before core collapse, and core silicon burning starts at around $x=-3$, 8 hours before core collapse.  At the time of core oxygen ignition ($x=-1.9$), an extended convective carbon-burning shell forms, reaching from $10\,\mathrm{M}_\odot$ to $30\,\mathrm{M}_\odot$ in mass coordinate and lasting until core collapse.  At core helium depletion, the envelope undergoes another brief ($2^{\mathrm{nd}}$) dredge-up phase by about $2\,\mathrm{M}_\odot$ and becomes a red supergiant with an extended convective envelope, from a mass coordinate of $34\,\mathrm{M}_\odot$ (helium core size) to the surface, that lasts until core collapse. 
\newline
\textsl{Right panel} A zoom-in of core and shell oxygen and silicon burning.  We use the same $x$-axis as in the left panel.  Starting just before $x=-3$ and at mass coordinate $3\,\mathrm{M}_\odot$ we see many small vertical stripes in the oxygen-burning shell.  These are due to an oscillatory instability occurring in this star so close to pair instability.  These oscillations encompass the entire core, and are hence also seen in the core and shell silicon burning.  In core silicon burning ($x=-3$ to $x=-3.3$) the oscillations also affect the convection, and, as before, the extended green regions are just the outlines of the many vertical convective zone boundaries.   Due to the logarithmic nature of the $x$-axis the oscillations appear to become wider toward the right-hand side of the plot although the frequency remains about constant, just the dynamical time-scale of the core.  A detailed frequency analysis is shown in Figure~\ref{fig:z85_pow}.
}
\label{fig:z85_khd}
\end{figure*}

\begin{figure}
\centering
\includegraphics[width=\columnwidth]{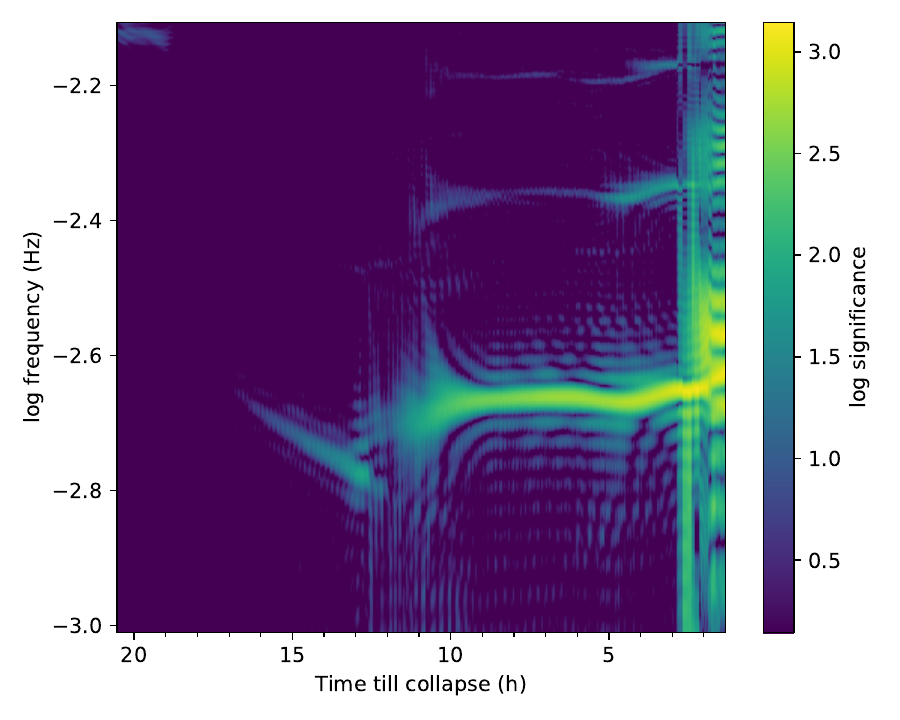}
\caption{Dynamical power spectrogram of the last $20$ hours of progenitor model \texttt{z85} prior to collapse using a time window of $2^{13}\,\mathrm{s}$.  Details on the spectrogram code used are described in \citet{2021MNRAS.500...34T}.  Here we plot the stellar neutrino luminosity of the star as this is an integral quantity over the entire star with a strong signal due to the high temperature-sensitivity of neutrino loss rates.  The onset of the clear signal with a frequency of about $10^{-2.65} \, \mathrm{Hz}$ (period of $450\,\mathrm{s}$) at $10\,\mathrm{h}$ before collapse corresponds to the clear oscillations seen starting in the Si shell at $3\,\mathrm{M}_\odot$ and $0.001\,\mathrm{yr}$ on the right-hand side of Figure~\ref{fig:z85_khd}.  This pulsational instability may be present earlier than seen in this model calculation but the simulation time step may have been too large to trace it in earlier evolution phases.}
\label{fig:z85_pow}
\end{figure}

\begin{figure*}
\centering
\includegraphics[width=\columnwidth]{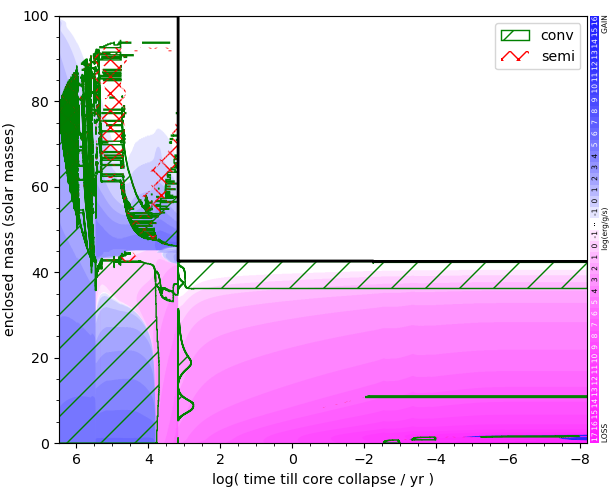}
\hfill
\includegraphics[width=\columnwidth]{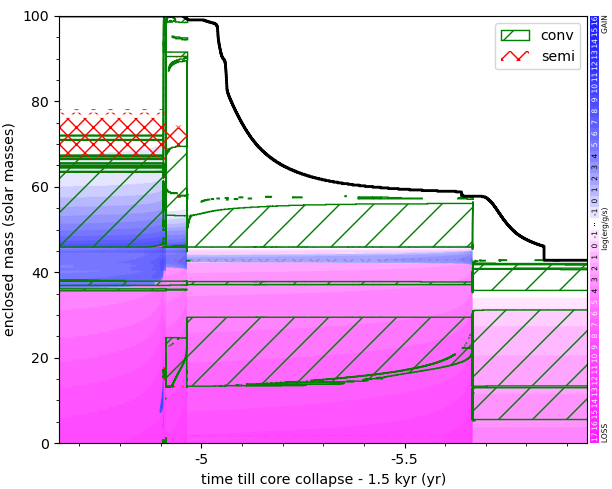}
\caption{Kippenhahn diagrams for the evolution of model \texttt{z100}.  
\newline
\textsl{Left panel}: The entire evolution from ZAMS to core collapse similar to Figure~\ref{fig:z85_khd}.   Core hydrogen burning ends at about $x=5.5$ and core helium burning ends at $x\approx 3.8$.  At $1.5\,\mathrm{kyr}$ prior to final collapse (x=3.176) the star undergoes a sequence of four pair-instability pulses (right panel, discussed below) that lead to the ejection of the hydrogen envelope, the helium shell, and the outer fringes of the CO core such that an oxygen-dominated core of only $42.53\,\mathrm{M}_\odot$ remains at the time of core collapse.  The outer convection zone seen here is in that CO core with a carbon mass fraction of only $5\,\%$.  In the final post-pulse evolution, core silicon burning occurs from $x=-2.5$ to $x=-3$ and two silicon-burning shells from $x=-3.2$ to $x=-4$ and $x=-5.2$ to $x=-8$.  Core neon and oxygen have already been depleted in powering the pair-instability pulses.
\newline
\textsl{Right panel}: 
At $\mathord\sim1\mathord,000\,\mathrm{yr}$ before the final core collapse, the star encounters radiative core and shell carbon burning and the pair instability sets in, leading to rapid contraction that is eventually stopped by ``explosive'' (rapid radiative) core neon and oxygen burning at $x=-4.9$ ($1\mathord,495.1\,\textrm{yr}$ prior to final core collapse).  Due to the pulse contraction only taking minutes, it is not clearly visible at the scale shown here for the purpose of providing an overview.  This is followed by a sequence of three further pulses on a recurrence time-scale of days to months.  See Figure~\ref{fig:z100_pulse} for details of the core density evolution.  Each of the pulses leads to rapid burning -- seen as shells rapidly burning outward in the lead-up to the pulse -- increasing entropy in the core and thereby reducing post-pulse density and temperature, visibly leading to a reduction in specific neutrino loss rates.  The star has to cool on the Kelvin Helmholtz time scale for the next pulse.  For the first pulses the cooling is clearly powered by neutrino losses, but after the last pulse neutrino losses become inefficient at first (see also \citealt{2017ApJ...836..244W}), leading to a longer recovery time to the final collapse with a quite altered core structure.  During the pulses, neon and oxygen are depleted in the core such that the usual hydrostatic convective neon and oxygen burning core and shell phases years to weeks prior to collapse as seen, e.g., for model \texttt{z85} (Figure~\ref{fig:z85_khd}; \citealt{2002RvMP...74.1015W}) cannot occur. 
}
\label{fig:z100_khd}
\end{figure*}

\begin{figure}
\centering
\includegraphics[width=\columnwidth]{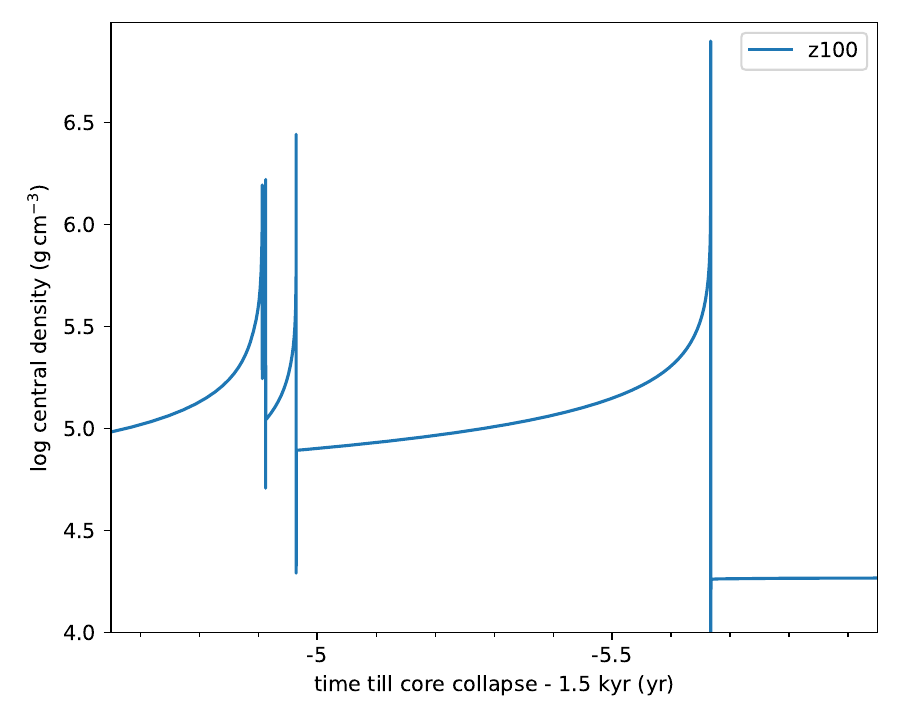}
\caption{Central density as a function of time of model \texttt{z100} for the same time range as in the right panel of Figure~\ref{fig:z100_khd}.  Shown are all four pulses of pulsational pair instability supernova (PPSN) the model encounters, $1.5\,\mathrm{kyr}$ before the final collapse.  The first two pulses are only 2 days apart, the third pulse occurs 19 days after the second, and the final fourth pulse occurs 145 days after the third pulse.  The pulses become increasingly more powerful, leading to lower post-pulse density with accordingly longer wait times for the next pulse as well as large mass ejections.  The maximum density peaks during the pulse ($\mathord{\sim}30\,\mathrm{s}$)  as well as the post-pulse ring-down (oscillation period is of a fraction of an hour and a comparable decay timescale) are not resolved in this plot and are visible as vertical lines only.  They are not relevant for the discussion of the presupernova models.  For more details on PPSNe see \citet{2017ApJ...836..244W}.}
\label{fig:z100_pulse}
\end{figure}

We simulate the collapse of two zero-metallicity (Pop-III) progenitor models, \texttt{z85} and \texttt{z100} \citep{heger_10}, with zero-age main sequence masses of $85\,\mathrm{M}_\odot$ and $100 \,\mathrm{M}_\odot$, respectively.  Both models are located close to the lower boundary of the pulsational-pair instability regime.  The progenitor models have been evolved up to collapse using the stellar evolution code \textsc{Kepler} \citep{weaver_78,2002ApJ...576..323R}.  Core density and entropy
profiles (Figure~\ref{fig:profiles})
and radial velocity profiles (Figure~\ref{fig:collapse})
of the two progenitor models reveal substantial structural differences.  In particular, model \texttt{z85} largely follows a typical massive star evolution path (Figure~\ref{fig:z85_khd}; \citealt{2002RvMP...74.1015W}) but encounters oscillatory-unstable oxygen shell burning (Figure~\ref{fig:z85_pow}) and eventually pair instability during iron core collapse (Figure~\ref{fig:collapse}), whereas the final structure of model \texttt{z100} is heavily affected by pair-instability pulses long before the final collapse (Figures~\ref{fig:z100_khd} and \ref{fig:z100_pulse}).

\subsection{Progenitor Model \texttt{z85}}
At first glance, the $85\,\mathrm{M}_\odot$ model exhibits a classical structure typical of most CCSN progenitors, just with a very massive core.  The low-entropy core inside the first convective shell has as mass of $2.43\, \mathrm{M}_\odot$, and both the jumps in specific entropy and density between the core and the surrounding shell are extremely well pronounced, with a huge shell specific entropy of $7.3\,k_\mathrm{B}\,\mathrm{u}^{-1}$, where $k_\mathrm{B}$ is the Boltzmann constant and $\mathrm{u}$ is the atomic mass unit. 
Both the Ertl criterion \citep{ertl_15} and the compactness criterion \citep{oconnor_11} firmly predict BH formation for this model due to the rather extreme values of the structural parameters $M_4=2.43$ and $\mu_4=2.90$ and a very high compactness of $\xi_{2.5}=0.86$.

A closer examination of the evolution of the model and its structure and composition at collapse reveal noticeable differences from normal CCSN progenitors.  Silicon core burning proceeds while oxygen burning above continues almost unaltered due to the large $\mathord\sim2.5\,\mathrm{M}_\odot$ silicon core.
At onset of core collapse, defined as the first model in which the infall velocity exceeds $900\,\mathrm{km}\,\mathrm{s}^{-1}$, the burning shell outside the core, from mass coordinate of $m=2.43\,\mathrm{M}_\odot$ to $m=6.77\,\mathrm{M}_\odot$, lives in the ashes of the previous oxygen-burning shell and is a violent silicon-burning shell with oxygen entrainment from the top, and the shell is in quasi-statistical equilibrium (QSE).  At the bottom of the shell, the mass fraction of iron group elements made by silicon burning reaches $85\,\%$ but drops to less than $1\,\%$ at the top of the shell.  In the outer region of the shell, clearly, the mixing time is longer than the burning time at the bottom.  The high entropy (Figure~\ref{fig:profiles}, orange dashed line) may be due to the oxygen entrainment; most of the oxygen does not reach the bottom of the shell but burns at a mass coordinate of $\mathord\sim4.5\,\mathrm{M}_\odot$, reflected in a local maximum in specific energy generation rate.   The actual oxygen burning shell starts at a mass coordinate of $m\approx 6.81\,\mathrm{M}_\odot$, separated from the silicon-burning shell by a semiconvective layer.  It is still a quite powerful shell with a specific energy generation rate that is comparable to that of the silicon-burning shell, and it has a high specific entropy of $8.2\,k_\mathrm{B}\,\mathrm{u}^{-1}$.

Concurrently with the collapse of the iron core, the entire outer part of the He core of $34.4\,\mathrm{M}_\odot$ is already collapsing with velocities of several $100\,\mathrm{km}\,\mathrm{s}^{-1}$ (Figure~\ref{fig:collapse}).  In effect, the model experiences a combination of concurrent ``classical'' core collapse and pair instability in the oxygen-rich shells.  The entire evolution of this model is shown in the Kippenhahn diagram in the left panel of Figure~\ref{fig:z85_khd}.  

A peculiarity of the model is its close proximity to the pair-instability regime.  As is not untypical for the transition between stable and unstable regimes (e.g., see \citealt{1983ApJ...264..282P,2007ApJ...665.1311H} for the case of accreting neutron stars), in this case we observe an oscillatory instability in oxygen shell burning and beyond.   In the right panel of Figure~\ref{fig:z85_khd} these oscillations become visible as horizontal stripes in the energy generation and neutrino loss rates -- both are very sensitive to temperature -- during the late part of the first oxygen shell burning and in silicon core and shell burning.  In fact, the oscillations may be present even at earlier times, however, the time step may have been too large to track them and the implicit hydro code would have smoothed them out.  A dynamic spectrogram for the neutrino signal of the oscillations is shown in Figure~\ref{fig:z85_pow}.  The oscillation starts (at least) $10\,\mathrm{h}$ before the collapse and has a frequency of about $2.2\,\mathrm{mHz}$.  As core collapse is reached, however, the global stability criterion \begin{equation*}
\int_0^M\frac{P}\rho\,\left(\gamma_\mathrm{ad}-\frac43\right)\,\mathrm{d} m > 0\;,
\end{equation*}
is violated and the inner $100\mathord{,}000\,\mathrm{km}$ undergo homologous collapse, superimposed with the homologous collapse of the iron core.  In the above equation, $P$ is the pressure, $m$ is the mass coordinate, $M$ is the total mass of the star, and $\gamma_\mathrm{ad}$ is the adiabatic index.   The oscillations slowly dampen out in the last half hour of contraction to the final core collapse, and the oscillatory instability transitions to a runaway growing instability.

%%%%%%%%%%%%%%%%%%%%%%%%%%%%%
\subsection{Progenitor Model \texttt{z100}}
The $100\,\mathrm{M}_\odot$ model has a distinctly different pre-collapse structure (Figure~\ref{fig:profiles}, blue lines).  Compared to Model \texttt{z85}, it has less extreme values of core mass and explodability parameters, with an Fe core mass of $1.69\,\mathrm{M}_\odot$, Ertl parameters
$M_4=2.06\,\mathrm{M}_\odot$, and $\mu_4=6.35$, and a compactness of $\xi_{2.5}=0.40$.  The Ertl criterion still indicates BH formation, however.  The shells outside the Fe core are non-convective, and there are no entropy and density steps associated with convective shell interfaces.
%$xi_{1.75}=0.8$

The unusual progenitor structure -- compared to normal CCSN progenitors -- is due to  the earlier evolution of the Model (Figure~\ref{fig:z100_khd}, left panel):
About $1\mathord,500$ years prior to collapse, the star experiences a sequence of four pair instability pulses of increasing strength and recurrence times (see \citealt{2017ApJ...836..244W}) within a few months.  These pulses eject the entire hydrogen envelope, the helium layer, and the outer part of the CO core, leaving behind a $42.53\,\mathrm{M}_\odot$ helium-free core that is dominated by oxygen, and with only small mass fractions of neon ($8\,\%$), carbon ($5\,\%$), and magnesium ($1\,\%$).  The pulses (Figure~\ref{fig:z100_khd}, right panel) lead to an increase of entropy and density (Figure~\ref{fig:z100_pulse}) in the core, and ``explosive'' oxygen and silicon burning during the pulses -- that actually powered the pulses -- lead to depletion of oxygen and even silicon in the core.  After the last pulse only a small amount of oxygen remained below $\mathord\sim5.5\,\mathrm{M}_\odot$, and in the centre a mass fraction of $66\,\%$ of iron group elements was made and only a mass fraction of $27\,\%$ of silicon and sulphur remained (plus $\mathord\sim7\,\%$ of calcium and argon).  As a result, in the final pre-collapse evolution there is no convective oxygen burning, neither core nor shell, and the silicon core and shell burning are rather weak and not very extended.  

It is this rather fast final evolution with runaway cooling and burning only in the centre, in the wake of the pair-instability pulse, that causes the rather high-entropy and low-density envelope: there is not enough time to lose entropy by neutrino emission after the pulse.  On close inspection, the difference can be seen as much more intensive purple shades before the first pulse at $x=-4.9$, in the right panel of Figure~\ref{fig:z100_khd}, as compared to the final distribution of purple shading in the core before collapse at $x=-8.2$ in the left panel of Figure~\ref{fig:z100_khd}.

%%%%%%%%%%%%%%%%%%%%%%%%%%%%%%%%%%%%%%%%%%%%%%%%%%
%%%%%%%%%%%%%%%%%%%%%%%%%%%%%%%%%%%%%%%%%%%%%%%%%%
\section{Supernova Simulations -- Numerical Methods and Setup}
\label{sec:simulation}
The simulations in this study are performed using the neutrino hydrodynamics code \textsc{CoCoNuT-FMT}. The setup of our 3D simulations is similar to our previous studies \citep{2019MNRAS.487.1178P, 2020MNRAS.494.4665P} with the exception of the EoS, however we repeat some of the details here for completeness.

We use a general relativistic finite-volume solver for the equations of hydrodynamics \citep{2002A&A...393..523D,2010ApJS..189..104M,mueller_19a}  formulated in spherical polar coordinates and the fast multi-group transport (FMT) method of \citet{mueller_15a} for the neutrino transport.
The GW emission is extracted by the time-integrated quadrupole formula \citep{finn_89,finn_90,blanchet_90} with relativistic correction factors as derived in \citet{mueller_13}.
The simulations are run with a spatial resolution of $550\times 128 \times 256$ zones in radius, latitude, and longitude. We employ a non-equidistant radial grid that reaches out to a radius of $10^5\, \mathrm{km}$. 

We use three different EoS at high densities that all match well with recent neutron star observations, namely the Lattimer \& Swesty EoS with a bulk incompressibility of K=220\,MeV (LS220; \citealp{Lattimer:1991nc}), and the SFHo and SFHx EoS from \citet{2013ApJ...774...17S}. For cold matter in $\beta$-equilibrium, the radius of a $1.4\,\mathrm{M}_\odot$ neutron star is 11.88\,km for SFHo, 11.97\,km for SFHx, and 12.62\,km for LS220. The maximum neutron star mass is $2.059\,\mathrm{M}_\odot$ for SFHo, $2.130\,\mathrm{M}_\odot$ for SFHx, and $2.06\,\mathrm{M}_\odot$ for LS220. 
These values are consistent with the latest constraints from GW observations \citep{PhysRevLett.121.161101, 2020NatAs...4..625C}, and pulsar and X-ray surveys \citep{PhysRevD.101.123007, Raaijmakers_2020}.
The SFHo and SFHx EoS are consistent with the latest nuclear constraints, however LS220 is incompatible with known nuclear constraints \citep{2017ApJ...848..105T}.
It should be pointed out, however, that compliance with constraints on cold neutron stars, the nuclear incompressibility, and the nuclear symmetry energy and its derivative 
does not necessarily guarantee (superior) accuracy in the supernova problem because of finite temperatures. Other parameters such as the nucleon effective mass can become critical \citep{yasin_20}; and arguments can be made that tuning the parameters of Skyrme-type or meson-exchange models to nuclear properties at saturation densities
is not sufficient to ensure correct behaviour in the supernova regime \citep{furusawa_17}. Given the remaining uncertainties about the EoS in the regime relevant to supernovae, an exploration of different models remains useful.
At low density, we use an EoS accounting for photons, electrons, positrons, and an ideal gas of nuclei together with a flashing treatment for nuclear reactions \citep{rampp_02}. 

In total, five different models have been simulated. Model \texttt{z85} has been simulated with all three EoS, and model \texttt{z100} has been simulated using the SFHx and SFHo EoS only.  The models are labelled as \texttt{PROGENITOR\_EoS}
(see Table~\ref{tab:properties}).

%%%%%%%%%%%%%%%%%%%%%%%%%%%%%%%%%%%%%%%%%%%%%%%%%%
%%%%%%%%%%%%%%%%%%%%%%%%%%%%%%%%%%%%%%%%%%%%%%%%%%
\section{Explosion Model Dynamics}
\label{sec:explosion}

\begin{table*}
\begin{center}
\begin{tabular}{|l|l|l|c|c|c|c|c|} 
 \hline
Model & Progenitor & EoS & $t_\mathrm{sh}$ & $E_\mathrm{diag}$ & $M_\mathrm{rem}$ & $t_\mathrm{BH}$ & $R_\mathrm{Shock}$ \\ 
      &           &     & (s)             & ($10^{51}\,\mathrm{erg}$)             & ($\mathrm{M}_{\odot}$)    & (s)              & (km) \\ \hline
\texttt{z85\_SFHx} & \texttt{z85}   & SFHx  & 0.298 & 2.7  & 2.57 & 0.59 & 4,451 \\ \hline
\texttt{z85\_SFHo} &\texttt{z85}    & SFHo  & 0.207 & 1.25 & 2.44 & 0.36 & 2,103 \\ \hline
\texttt{z85\_LS220} &\texttt{z85}   & LS220 & 0.160 & 0.7  & 2.51 & 0.29 & 1,504 \\ \hline
\texttt{z100\_SFHx} & \texttt{z100} & SFHx  & ---   & ---         & 1.88 & $\gg 0.5$ & 89 \\ \hline
\texttt{z100\_SFHo} & \texttt{z100} & SFHo  &  ---  & ---         & 2.05 & $\gg 0.5$ & 60 \\ \hline
\end{tabular}
\caption{Summary of model setup and outcomes.
$t_\mathrm{sh}$  is the time of shock revival, $E_\mathrm{diag}$ is the diagnostic energy at the end of the simulation, $M_\mathrm{rem}$ is
the mass of remnant at the end of the simulation, $t_\mathrm{BH}$ is the time after bounce of black hole formation, and $R_\mathrm{shock}$ is the shock radius at the end of the simulation. The SFHx EoS, which supports the highest
maximum mass, is the least favourable to explosion but produces a larger diagnostic energy due to the longer accretion time. }
\label{tab:properties}
\end{center}
\end{table*}

\begin{figure*}
\centering
\includegraphics[width=0.99\columnwidth]{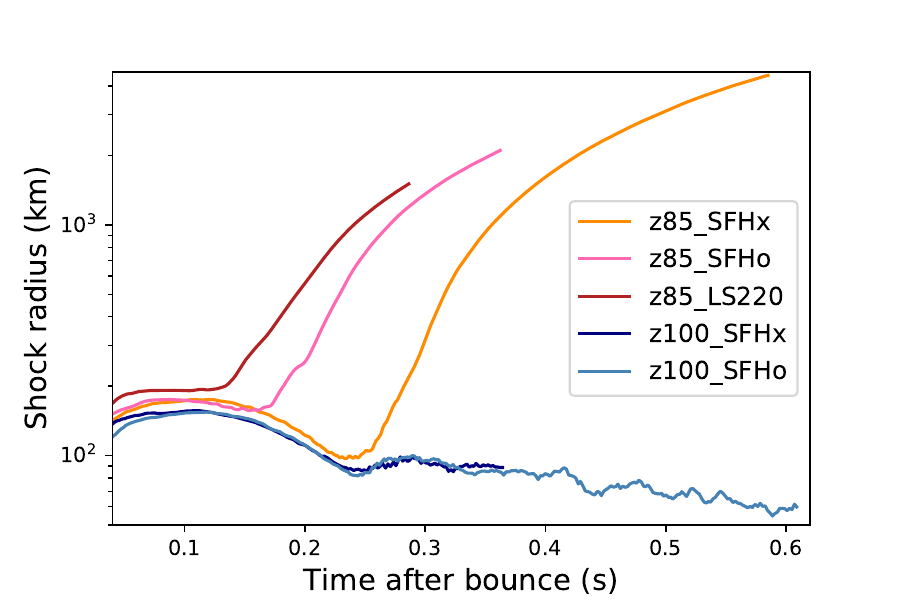}
\includegraphics[width=0.99\columnwidth]{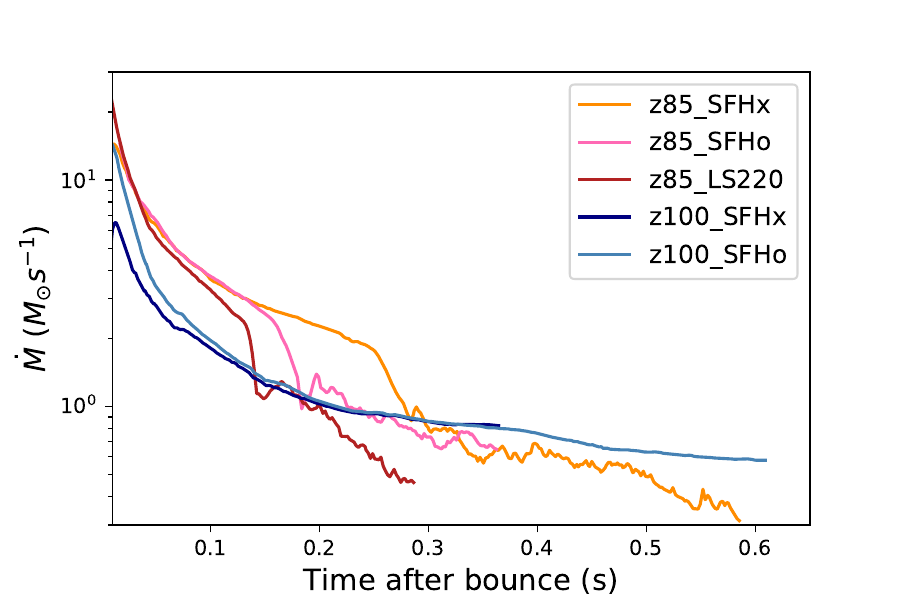}
\includegraphics[width=0.99\columnwidth]{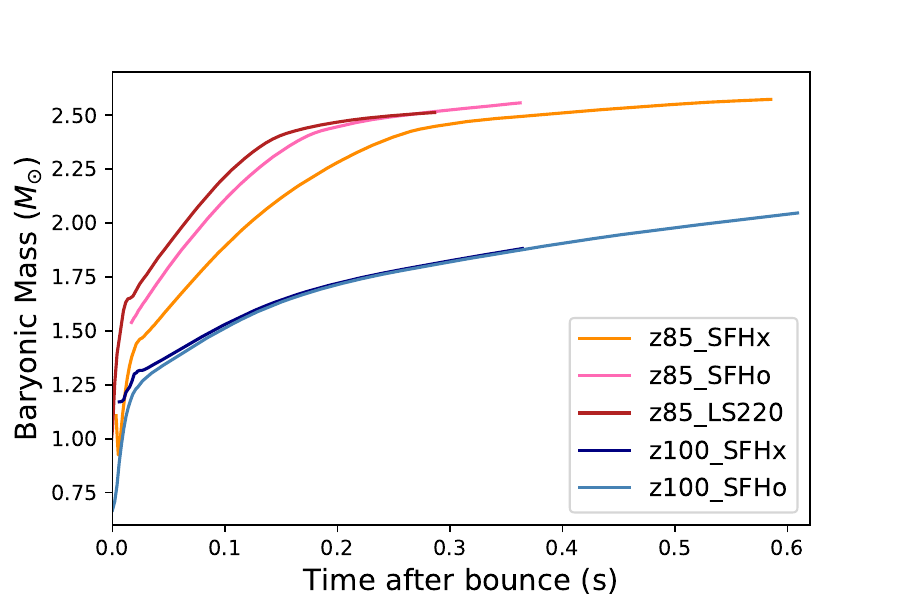}
\includegraphics[width=0.99\columnwidth]{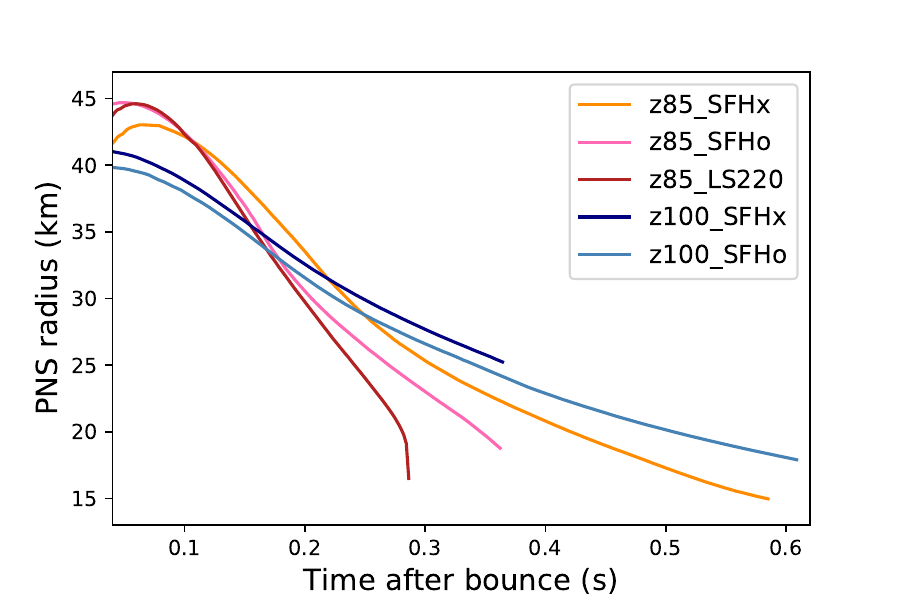}
\caption{Angle-averaged shock radii 
(top left), mass accretion rate at 200\,km (top right)
baryonic PNS mass (bottom left), and PNS radius (bottom right) for all models.  The \texttt{z85} models achieve shock revival early after core bounce and then form black holes within hundreds
of milliseconds. The \texttt{z100} models have a high mass accretion rate but do  not reach black hole formation before the end of the simulation.}
\label{fig:shock}
\end{figure*}

In this section, we discuss the dynamical evolution and, where applicable, the explosion and remnant properties of our models.
The outcomes of the five simulations are summarised in Table~\ref{tab:properties}.
The average shock radii for all models are shown in Figure~\ref{fig:shock} (top left panel). The three \texttt{z85} models all undergo shock revival before BH formation, whereas
the shock still has not started to move out in the \texttt{z100} models.
In some respects, the behaviour of the two progenitors corresponds
to trends found by \citet{2018ApJ...855L...3O} in that the \texttt{z85} models
with a very massive low-entropy core and high post-bounce mass accretion
rates (Figure~\ref{fig:shock}, top right panel) explode more readily
than the \texttt{z100} models with lower post-bounce accretion rates. A close
examination of the two sets of models reveals important differences
to the findings of \citet{2018ApJ...855L...3O}, however.

%%%%%%%%%%%%%%%%%%%%%%%%%%%%%%%%%%%%%%%%%%%%
\subsection{Progenitor Model \texttt{z85}}
\begin{figure}
\centering
\includegraphics[width=\columnwidth]{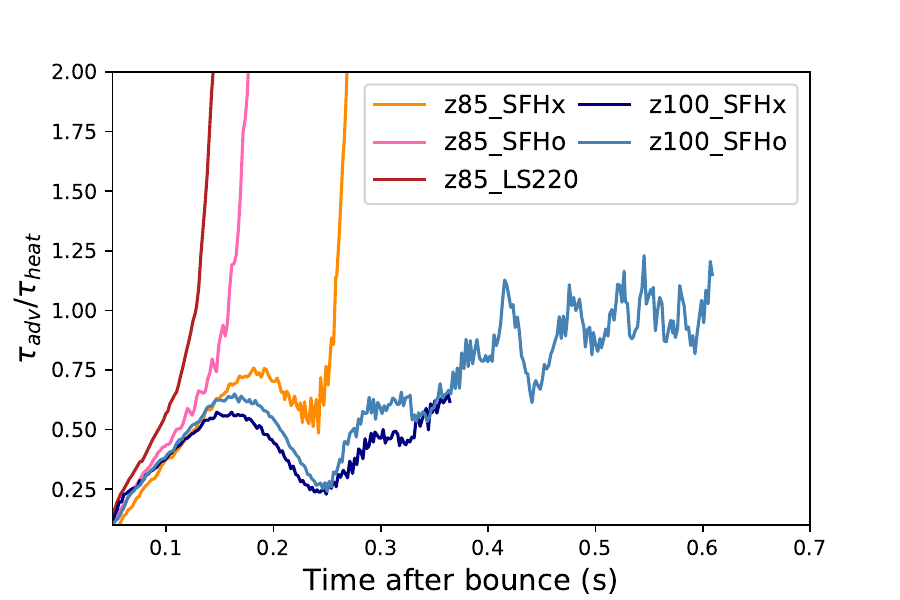}
\caption{The time-scale criterion $\tau_\mathrm{adv}/\tau_\mathrm{heat}$ for all models. The time-scale criterion indicates conditions for neutrino-driven runaway shock expansion at $\tau_\mathrm{adv}/\tau_\mathrm{heat} \gtrsim 1$. The time-scale criterion values for the \texttt{z100} models indicate that the shock would likely be revived for these models some time after 0.6\,s.  
}
\label{fig:criticality}
\end{figure}

In the \texttt{z85} explosion models, shock revival occurs early after bounce
with the average shock radius crossing 300\,km 
at
$0.160\, \mathrm{s}$ (LS220)
$0.207\, \mathrm{s}$ (SFHo),
and $0.298 \, \mathrm{s}$ (SFHx), respectively
(Table~\ref{tab:properties}).
The ratio between the advection and heating time scale
$\tau_\mathrm{adv}$ and $\tau_\mathrm{heat}$, which
quantifies the proximity to runaway shock expansion
\citep{janka_01,buras_06a,2020LRCA....6....3M}, 
exceeds the critical threshold $\tau_\mathrm{adv}/\tau_\mathrm{heat}=1$
even earlier at times of 
0.128\,s (LS220), 0.158\,s (SFHo), and 0.257\,s (SFHx)
(Figure~\ref{fig:criticality}) when the shock
is still within the low-entropy core as can be seen from
the PNS masses at the corresponding times
(Figure~\ref{fig:shock}, bottom left).
It is difficult to unambiguously associate the
different shock trajectories and proto-neutron star radii
(Figure~\ref{fig:shock}, bottom right) with the microphysical
properties of the different EoS.
For EoS that differ more markedly the impact of the microphysics
on the heating conditions is now better understood; for example
\citet{yasin_20} recently identified the low effective nucleon mass
as the critical factor that explains adverse heating conditions
in case of the Shen EoS \citep{shen_98}. While this may
play a role in explaining the different explosion times of the
\texttt{z85} models since the SFHo and SFHx EoS also have lower
effective nucleon masses than LS220  \citep{2013ApJ...774...17S},
there are confounding factors that might influence the order
of shock revival.
The models with different EoS exhibit differences
in mass accretion rate and PNS mass already early on.
This is due to the interplay of different collapse times
for the three EoS and the peculiar evolution of
the progenitor towards collapse with a pair instability
pulse that coincides with core collapse. An influence
of minute EoS differences on the collapse time
and early accretion history has been noted before
\citep[see][Section 3.2]{huedepohl_phd}, and is ideally avoided
 by using the same low/intermediate-density EoS
of different simulations up to densities of
$\mathord{\sim} 10^{11}\,\mathrm{g}\,\mathrm{cm}^{-3}$
\citep{huedepohl_phd,bruenn_20}.
It is also worth pointing out that $\tau_\mathrm{adv}/\tau_\mathrm{heat}$
 already reaches a value of $0.75$ in model \texttt{z85\_SFHx} before
$0.2\, \mathrm{s}$ after bounce and only narrowly fails to explode
earlier. The delay in shock revival compared
to \texttt{z85\_LS220} and \texttt{z85\_SFHo}
may thus give
an exaggerated difference between intrinsic EoS differences.
It remains to be determined how the robust the hierarchy of
shock revival times between LS220, SFHo, and SHFx is, but
we do note that the order of shock revival
between the LS220, SFHo, and SFHx models \emph{is} consistent with other
recent studies \citep{bollig_20,landfield_phd},

All three models form BHs soon after shock expansion
sets in. BH formation occurs at $0.59\,\mathrm{s}$, $0.36\,\mathrm{s}$ and $0.29\,\mathrm{s}$ after core bounce for the SFHx, SFHo and LS220 models, respectively.
Interestingly, even though model \texttt{z85\_SFHx} takes the longest time to reach shock revival, the shock has propagated further than in the other two
models by the time of BH collapse\footnote{We define the BH formation time as the point where the central density exceeds the boundaries of the EoS table. The boundary is encountered slightly earlier 
for the SFHx and SFHo EoS than for LS220. Since BH formation generally
occurs between the 2\,ms output intervals, there is usually no output file available exactly at this point in time. In the last output file,
the central density and lapse have typically reached values of $\rho_\mathrm{c}\gtrsim 10^{15}\,\mathrm{g}\,\mathrm{cm}^{-3}$ and
$\alpha_\mathrm{c}=0.4\texttt{-}0.45$, respectively.
} At least qualitatively, the differences in BH formation
time can be more easily explained than the differences in explosion time.
After about $250\, \mathrm{ms}$ the PNS masses have become quite similar
and the different maximum mass of warm neutron stars becomes the
most important factor \citep[cf.][]{2013ApJ...774...17S,da_silva_20} that results in 
\texttt{z85\_SFHx} forming a BH later than
\texttt{z85\_SFHo} and \texttt{z85\_LS220}.
The mass accreted onto the PNS is also higher for
\texttt{z85\_SFHx} ($0.114\,\mathrm{M}_\odot$)
than for  \texttt{z85\_SFHx} ($0.100\,\mathrm{M}_\odot$)
and \texttt{z85\_LS220} ($0.085\,\mathrm{M}_\odot)$.

Model \texttt{z85\_SFHx} also has
the highest diagnostic explosion energy at the time of BH collapse, with a value of $2.7\times 10^{51}\,\mathrm{erg}$
as opposed to
$1.25\times 10^{51}\,\mathrm{erg}$ for \texttt{z85\_SFHo} and \texttt{z85\_LS220}.
This is due to the longer accretion time before the maximum PNS mass is reached.  The energetics and ultimate fate (i.e., whether the
shock manages to propagate outward and expel the envelope)
of supernovae with early fallback could therefore prove very
sensitive to the nuclear EoS.

To determine the final fate of the ``aborted'' explosion in the
three $85\,\mathrm{M}_\odot$ models, long-time simulations in the vein
of \citet{2018ApJ...852L..19C} would be required. Despite progress
on the theory of mass ejection by weak explosions
\citep{chan_20,mandel_20,matzner_20,linial_20}, several scenarios
are conceivable.
In all cases, the binding energy of the shells outside the
shock by far exceeds the diagnostic explosion energy at the
time of BH collapse; even for model \texttt{z85\_SFHx}, this ``overburden''
is still $5\texttt{-}6\times 10^{51}\, \mathrm{erg}$.
Since, however, the pre-shock infall velocities
are already subsonic in model \texttt{z85\_SFHx}, it is likely that the shock
will continue to propagate outwards for a substantial time and transition
to the weak-shock regime as it scoops up bound pre-shock material 
\citep{2018ApJ...852L..19C,chan_20}. It has been argued
\citep{mandel_20,matzner_20,linial_20} that the energy or acoustic
luminosity of the resulting sound pulse is approximately conserved
and then determines the amount of material ejected from the surface.
Given the large ratio of envelope binding energy to diagnostic explosion
energy it is still a distinct possibility that the shock will not reach the surface. 

For model \texttt{z85\_LS220}, the situation is different in that the shock has just barely reached the sonic point of the infall region at the time of BH collapse, and the shock is already weaker to begin with. It therefore appears likely that the shock
cannot escape the newly formed BH since a weak sound pulse would be too slow to propagate outward through the infalling pre-shock matter.

Despite these uncertainties, we can obtain a conservative \emph{lower} limit for the final BH masses of our models. Using the (extremely optimistic) assumption that the sound pulse carries the initial explosion energy without any loses, we can match this energy with the binding energy of the ejected shells and thereby estimate the minimum final BH masses of $30.7\,\mathrm{M}_\odot$ for \texttt{z85\_SFHx}, $32.4\,\mathrm{M}_\odot$ for \texttt{z85\_SFHo}, and $33.2\,\mathrm{M}_\odot$ for \texttt{z85\_LS220}. In case the hydrogen envelope of the progenitor has been lost due to binary interaction, we would  predict a fairly narrow range of $30.7\texttt{-}34.4\,\mathrm{M}_\odot$ for the BH mass.

%%%%%%%%%%%%%%%%%%%%%%%%%%%%%%%%%%%%%%%%%%%%
\subsection{Progenitor Model \texttt{z100}}
The two \texttt{z100} explosion models do not achieve shock revival before the end of the simulation time. 
The time-scale criterion $\tau_\mathrm{adv}/\tau_\mathrm{heat}$ has a clear upward trend for these two models, however, and  has already reached the critical value $\tau_\mathrm{adv}/\tau_\mathrm{heat}=1$
at $0.41 \, \mathrm{s}$
for \texttt{z100\_SFHx} (Figure~\ref{fig:criticality}). 
It is therefore likely that the shock would
be revived in these two models some time
after $0.6\, \mathrm{s}$. Even though the mass accretion rate is still quite high in the \texttt{z100} model at the end of the simulations, the baryonic PNS masses are still quite far away from the maximum values allowed for their respective EoS. The most probable outcome for these models is therefore that they will experience shock revival, but still undergo delayed collapse to a BH. Since the binding of the shells ahead of the shock is $\gtrsim 4\times 10^{51}\, \mathrm{erg}$, it is unlikely that a neutrino-driven explosion could still become strong enough to completely expel the envelope. As we could not follow these two models into the explosion phase and up to BH formation, we cannot assess whether there is any chance of partial mass ejection, or whether the entire metal core left by the previous pair instability pulse will completely collapse to a BH.

%%%%%%%%%%%%%%%%%%%%%%%%%%%%%%%%%%%%%%%%%%%%%%%%%%
\subsection{EoS dependence of SASI activity}

\begin{figure*}
\centering
\includegraphics[width=\textwidth]{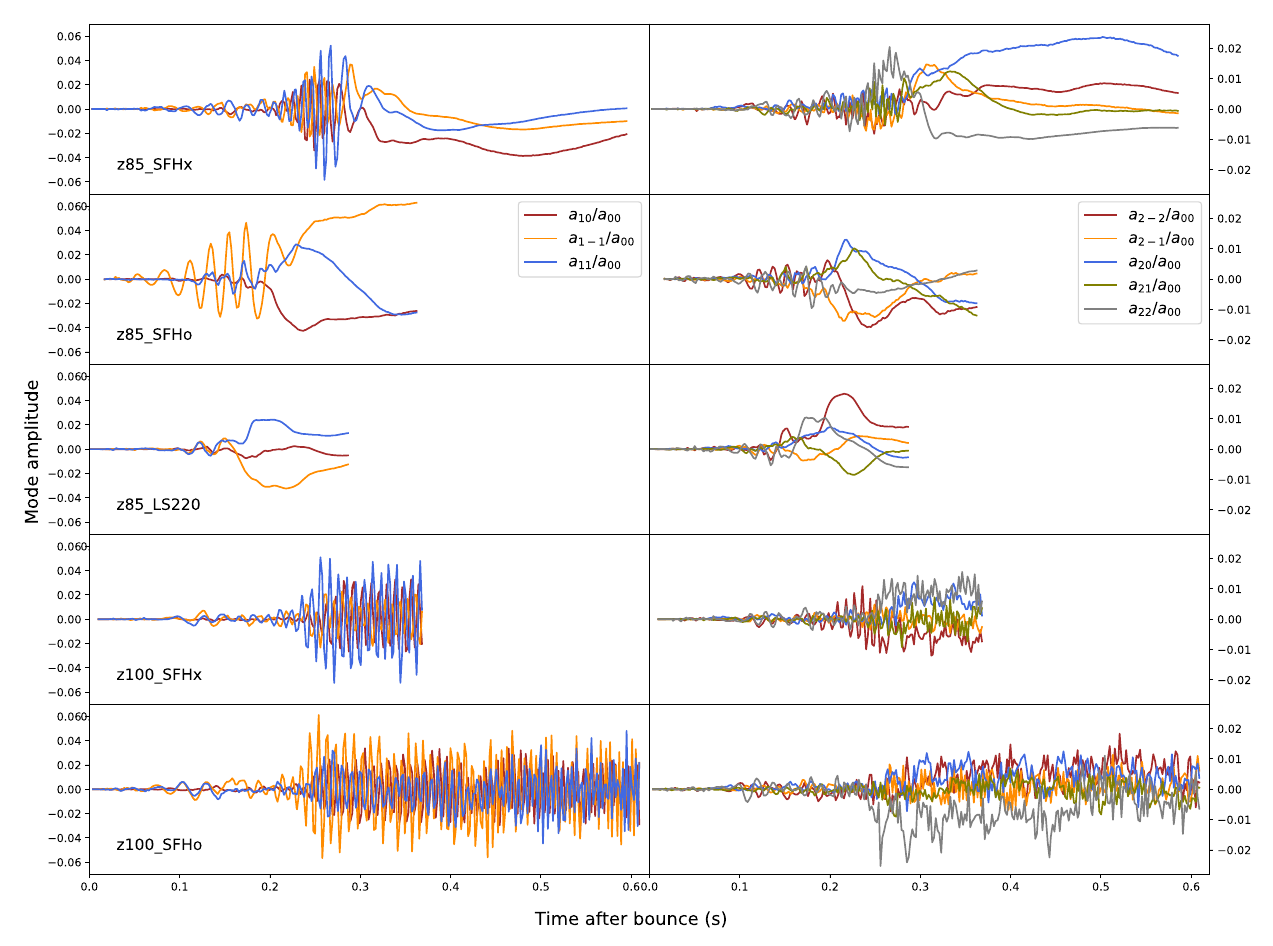}
\caption{Normalised dipole 
($a_{1m}/a_{00}$) and quadrupole coefficients 
($a_{2m}/a_{00}$) of the angle-dependent
shock position.}
\label{fig:coeffients}
\end{figure*}

\begin{figure*}
\centering
\includegraphics[width=0.90\columnwidth]{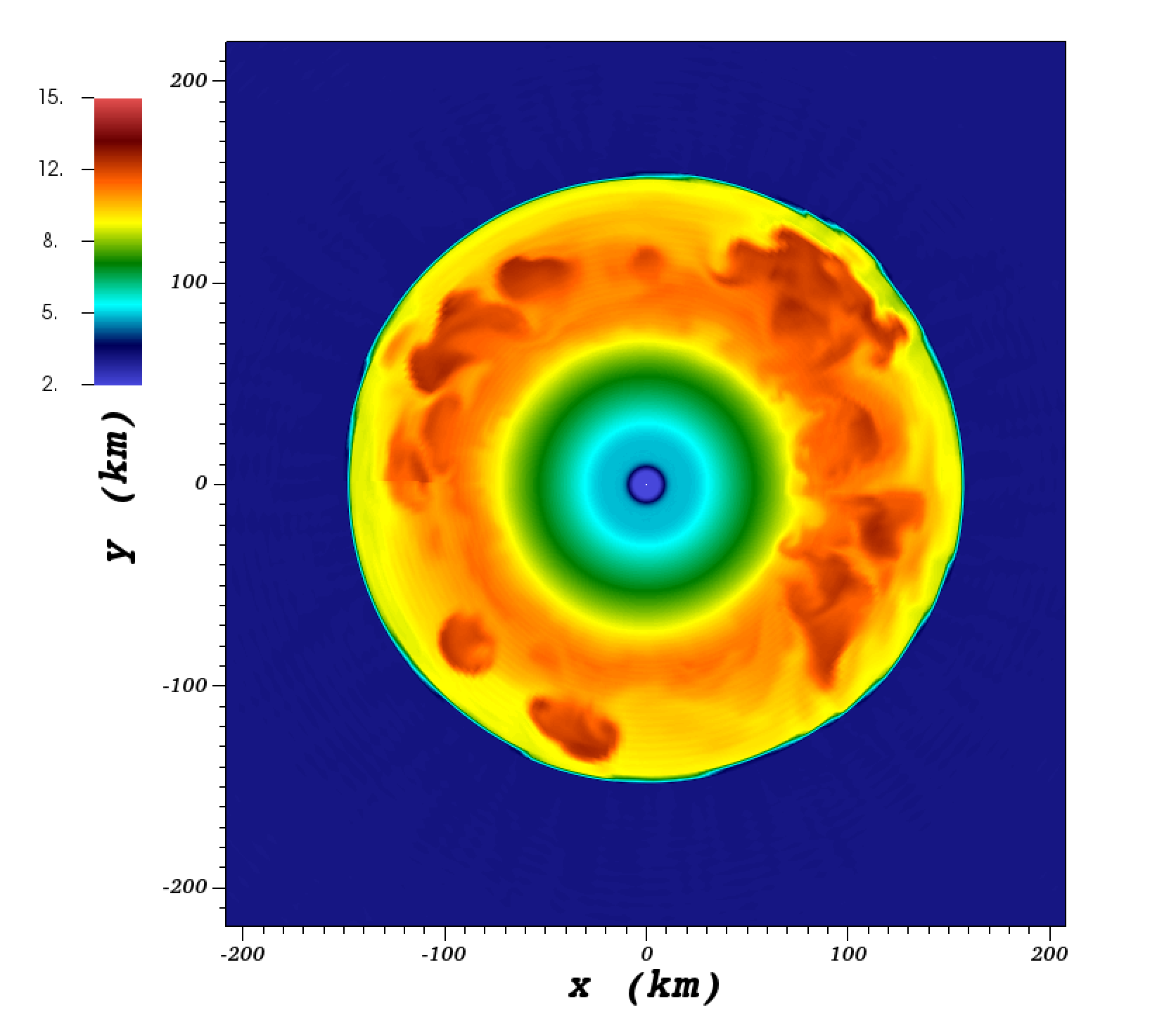}
\includegraphics[width=0.90\columnwidth]{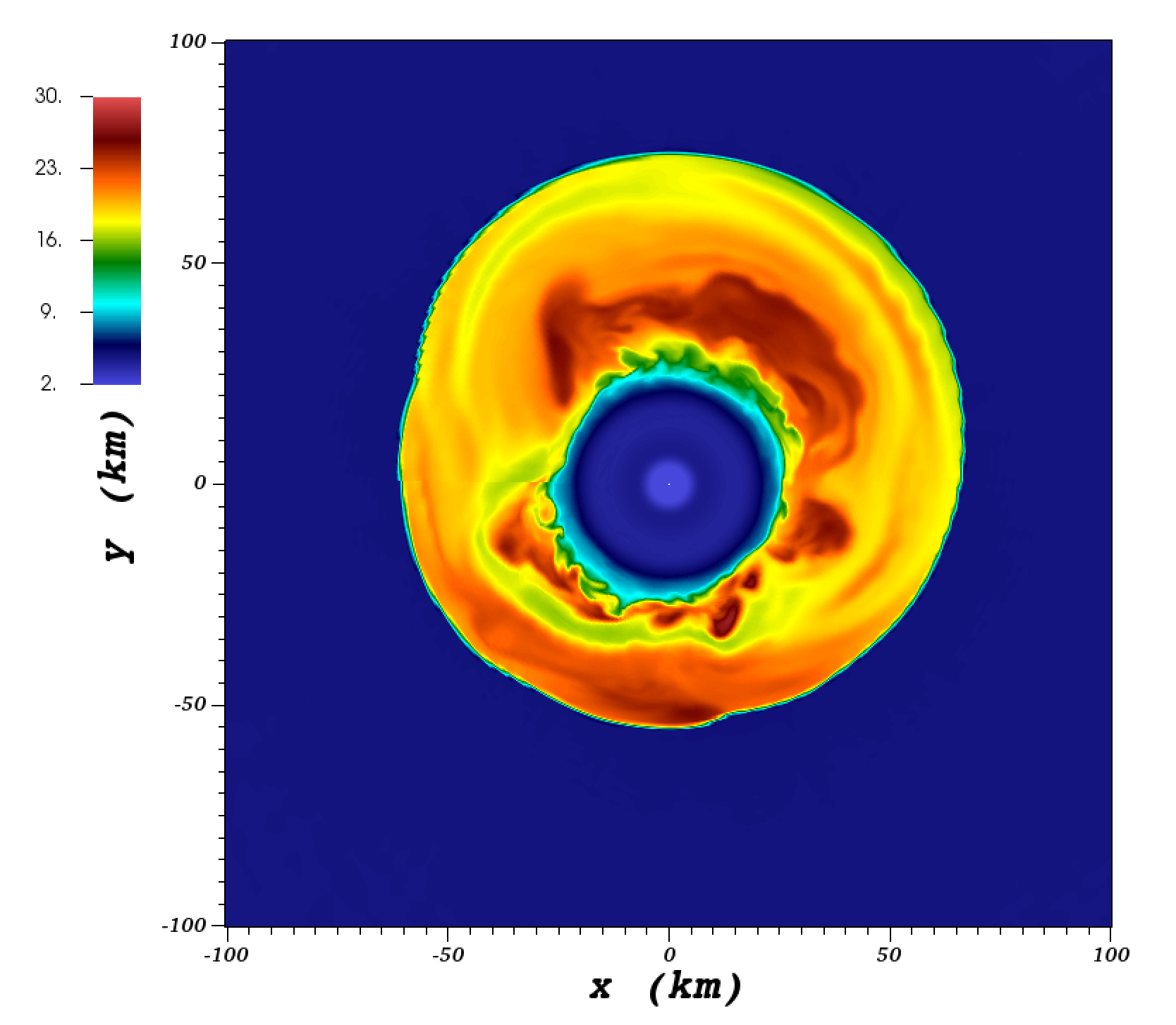}
\includegraphics[width=0.90\columnwidth]{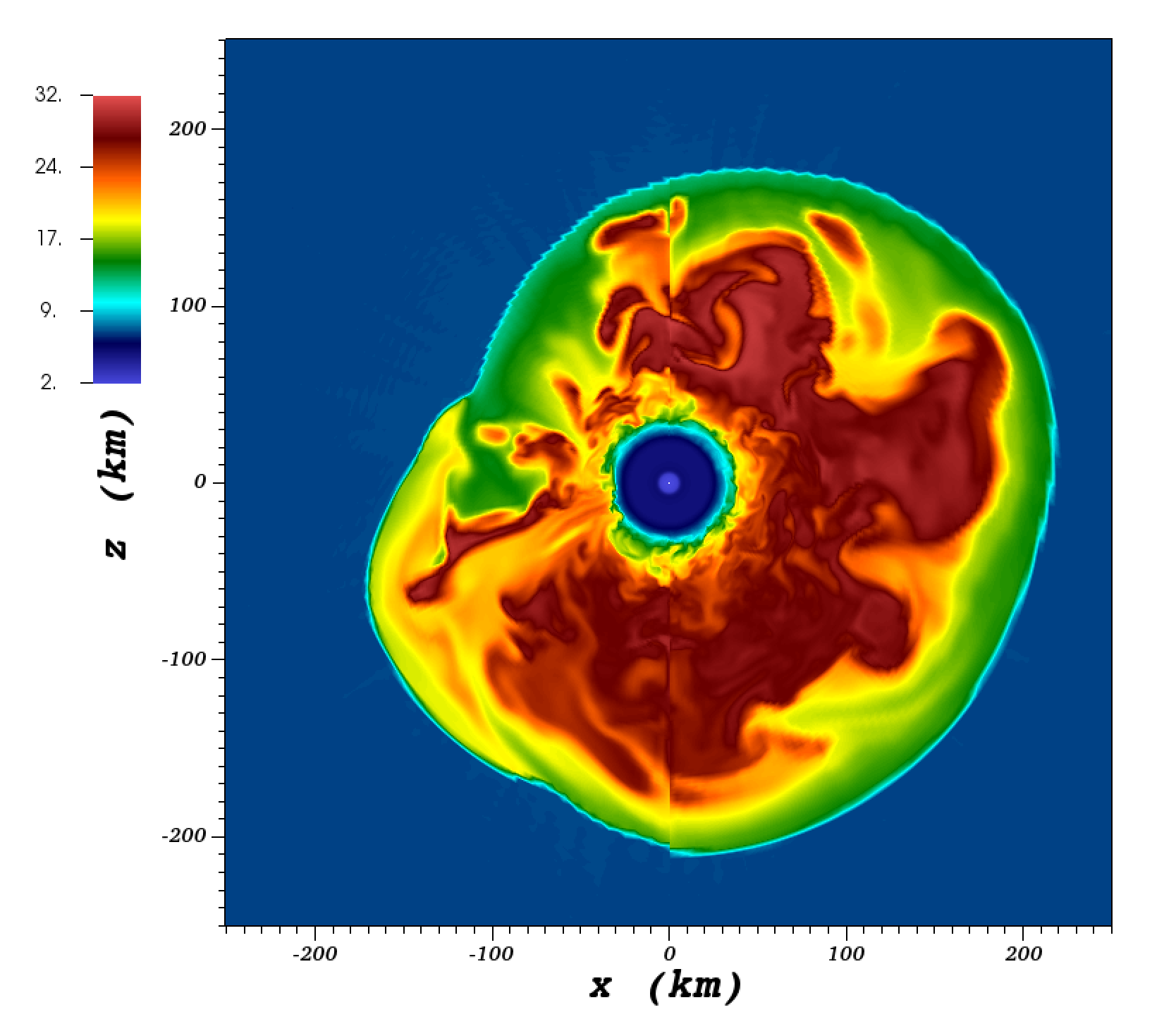}
\includegraphics[width=0.93\columnwidth]{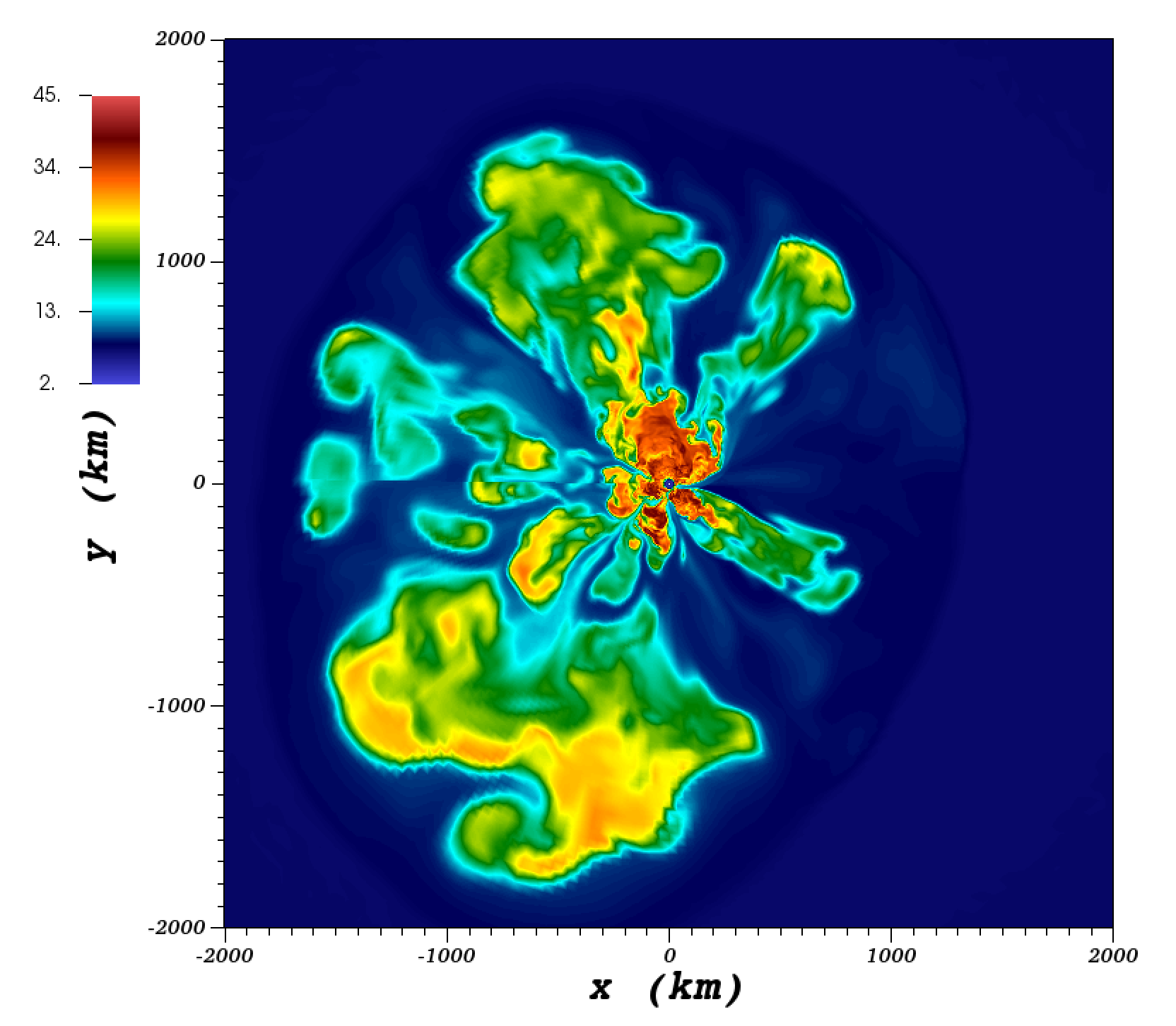}
\caption{2D slice of the entropy in model \texttt{z100\_SFHo} at 100\,ms and 450\,ms post-bounce (top)
and at 280\,ms and 450\,ms in model \texttt{z85\_SFHx}. All the SFHo and SFHx models develop strong SASI activity. In the \texttt{z100} models, SASI starts out as weak with some Rayleigh-Taylor convection on top (top left panel) and then becomes stronger with a more distinct SASI flow morphology later on (top right). In the \texttt{z85} models, strong SASI develops early on (bottom left) and plays a crucial role in expanding the shock radius until neutrino heating conditions become critical. The SASI then freezes out and buoyancy-driven outflows develop in the explosion phase (bottom right).
}
\label{fig:entropy}
\end{figure*}

We find that the EoS qualitatively affects the nature of the hydrodynamic instabilities during the pre-explosion phase. To diagnose SASI activity in our models, we decompose the angle-dependent shock position $r_\mathrm{sh}(\theta,\varphi)$ into spherical harmonics $a_{lm}$,
\begin{equation}
a_{lm}=
\frac{(-1)^m}{\sqrt{4\pi(2l+1)}}
\int Y^*_{lm}(\theta,\varphi) 
r_\mathrm{sh}(\theta,\varphi)\,\mathrm{d}\Omega,
\end{equation}
where $Y_{lm}$ are real spherical harmonics with the same normalisation as in \citet{burrows_12}. In Figure~\ref{fig:coeffients}, we show the normalised dipole and quadrupole coefficients of the shock position. We also illustrate the multi-dimensional structure of the flow in models \texttt{z85\_SFHo} and \texttt{z100\_SFHo} at selected epochs using 2D slices of the entropy in the supernova core in Figure~\ref{fig:entropy}.

All of the SFHo and SFHx models develop strong SASI activity at some point, whereas model \texttt{z85\_LS220} hardly develops quasi-periodic shock oscillations and is clearly convectively dominated around shock revival. 
A trend towards strong SASI activity with the SFHx EoS was already found by \citet{kuroda_16}. It is noteworthy that strong SASI also occurs in the exploding models \texttt{z85\_SFHx} and \texttt{z85\_SFHo} in contrast to the findings of \citet{2018ApJ...855L...3O}, who posited that rapidly developing explosions in progenitors with high compactness are dominated by convection from the outset. Models \texttt{z85\_SFHx} and \texttt{z85\_SFHo} rather develop SASI activity earlier than the \texttt{z100} models with lower accretion rates. The \texttt{z100} models rather go through a regime where rather weak SASI activity and convective plumes can be seen side by side (Figure~\ref{fig:entropy}, top left) before developing stronger and cleaner
SASI oscillation later from about $0.25\,\mathrm{s}$ onward (Figure~\ref{fig:entropy}, top right). The SASI then maintains strong and stable dipole modes (for several hundred milliseconds in \texttt{z100\_SFHo}).

In all models with SASI activity, the dipole mode appears to be dominant. The models do not show pronounced quasi-periodic oscillations in the quadrupole coefficient $a_{2m}$ most of the time. There are, however, hints of modest quasi-periodic quadrupolar oscillations in  model \texttt{z85\_SFHo} between $0.1\, \mathrm{s}$
and $0.2\, \mathrm{s}$ and in \texttt{z100\_SFHx} between  $0.2 \, \mathrm{s}$ and $0.25 \, \mathrm{s}$. Similar to the BH-forming
models of \citet{2020PhRvD.101l3013W},
a pronounced SASI quadrupole only appears
episodically, and different from the models
of \citet{2020PhRvD.101l3013W}. This does not, of course, \emph{not} argue against the existence of a regime with a dominant quadrupole mode in some BH-forming progenitors; a dominant quadrupole simply does not emerge for the two particular progenitors considered in this study, and the emergence of a dominant quadrupole may hinge on details of the neutrino transport and EoS effects like muonisation 
\citep{bollig_17} that are not included in our models.

%%%%%%%%%%%%%%%%%%%%%%%%%%%%%%%%%%%%%%%%%%%%%%%%%%%%%%%%%
%%%%%%%%%%%%%%%%%%%%%%%%%%%%%%%%%%%%%%%%%%%%%%%%%%%%%%%%%
\section{Gravitational Waves}
\label{sec:gravwaves}

\begin{figure*}
\centering
\includegraphics[width=\textwidth]{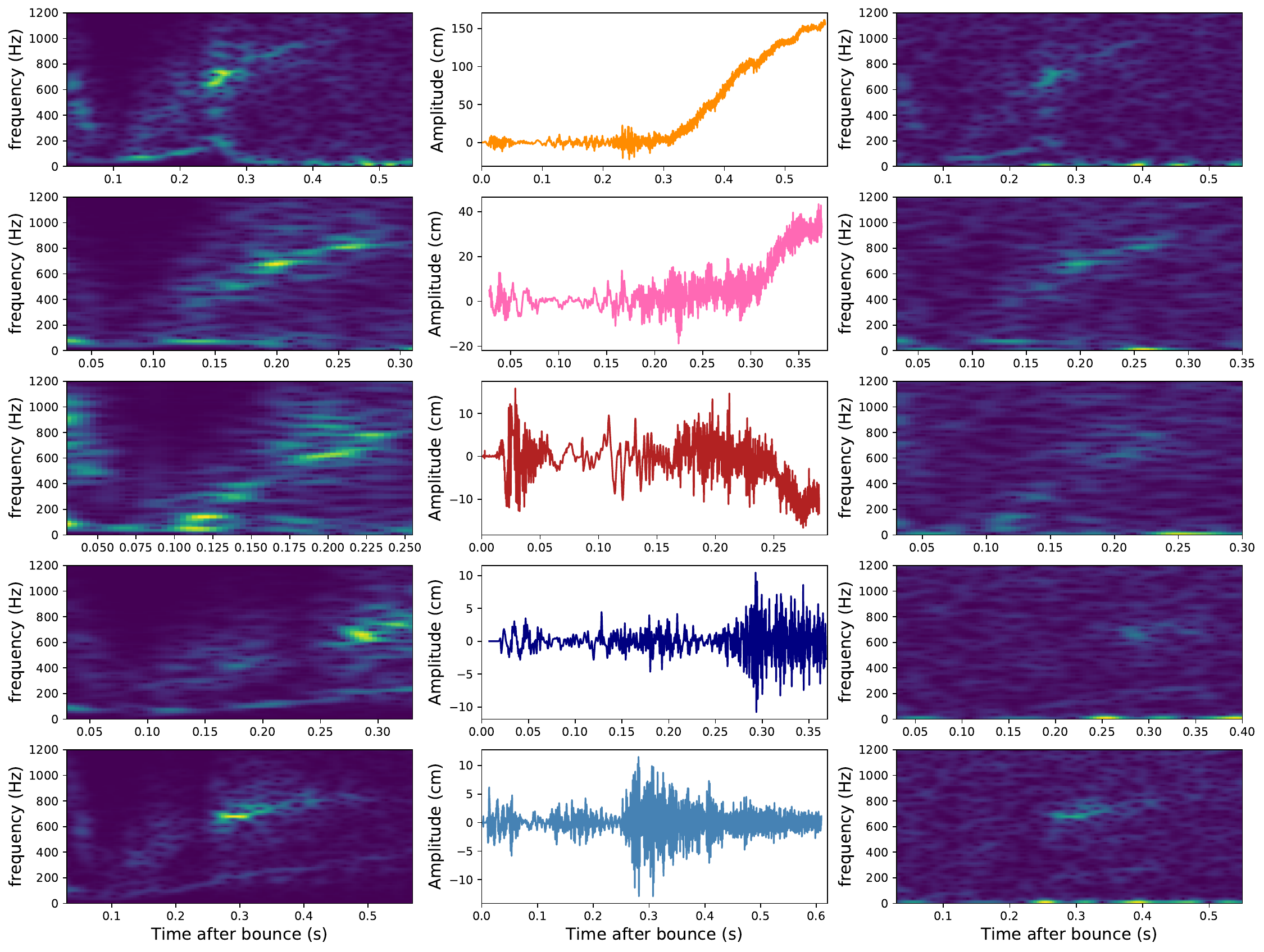}
\caption{Gravitational wave signals
and spectrograms for models
\texttt{z85\_SFHx}, \texttt{z85\_SFHo}, \texttt{z85\_LS220}, \texttt{z100\_SFHx}, and \texttt{z100\_SFHo} (top to bottom).
The middle column shows the amplitude $h_+$ of the plus polarisation mode for an observer
in the equatorial plane at
$(\theta,\varphi)=(90^\circ,90^\circ)$. The left and right columns show
the corresponding spectrograms without noise (left)
and with Gaussian design sensitivity aLIGO noise at a distance of 5\,kpc (right).}
\label{fig:spectrograms}
\end{figure*}

\subsection{Features of the GW signal}
The time series and spectrograms of the GW emission of our models are shown in Figure~\ref{fig:spectrograms} for a selected observer direction in the equatorial plane
of the spherical polar grid. All of the models show the typical g-mode emission which rises in frequency with time from a few hundred Hz to $\sim 900$\,Hz.  
The effects of the different EoS are imprinted on the g-mode GW signals.
As the g-mode GW frequency is $\propto M_\mathrm{PNS}/R_\mathrm{PNS}^2$ \citep{mueller_13}, the different EoS change how quickly the g-mode frequency rises in time. The rise will be dictated by the effective warm mass-radius relation R(M) (defined by a density of $10^{11}\texttt{-}10^{12}\, \mathrm{g}\, \mathrm{cm}^{-3}$
examples of which are shown, e.g., in Figure~3 of \citealt{2017PhRvD..96f3005S}), which is related to the properties of the EoS and the time dependence of $M(t)$ . The time-dependence of $M(t)$ could be reconstructed from the neutrino signal \citep{2014ApJ...788...82M}. One should note, however, that the warm mass-radius relation does depend on the entropy profiles and is somewhat progenitor- and time-dependent. Precision measurements of EoS properties through the g-mode frequency are therefore not realistic. The GW frequency increases more rapidly for LS220 than for SFHo or the SFHx model which has the slowest rise in GW frequency with time. 
In the \texttt{z85} models, the GW power from this mode peaks shortly after shock revival, which is also affected by the different EoS. As a consequence, the frequency around peak emission (which will mostly determine the overall spectrum) is lower in models that explode faster, in our case $\mathord{\sim} 650\,\mathrm{Hz}$ for LS220, $\mathord{\sim} 695\,\mathrm{Hz}$ for SFHo, and $\mathord{\sim} 710\,\mathrm{Hz}$ for SFHx. 
The \texttt{z100} models, which do not achieve shock revival, reach their maximum GW amplitudes
between $0.25\,\mathrm{s}$ and $\mathord{\sim} 0.3\,\mathrm{s}$ once the shock has contracted
sufficiently for strong SASI to set in. As the shock contracts further, the mass in the gain region
decreases, the SASI motions that excite the g-mode carry less energy, again resulting in smaller GW amplitudes. Peak GW emission from the g-mode occurs at frequencies of $\mathord{\sim} 700\,\mathrm{Hz}$ for \texttt{z100\_SFHx} and $\mathord{\sim} 695\,\mathrm{Hz}$ for \texttt{z100\_SFHo}. 

The g-mode emission of all our models is of lower frequency than the results obtained in some recent work by other groups which still have high GW amplitudes at frequencies over $1,000\,\mathrm{Hz}$ \citep{2018ApJ...865...81O, 2019ApJ...876L...9R, 2020PhRvD.102b3027M, 2020arXiv201002453P}. This means that our models are in a better frequency band for current ground-based GW detectors, and those from other groups may be more promising sources for proposed future high frequency GW detectors \citep{2020arXiv200703128A}. It is also noteworthy that BH-forming models will not \emph{generically}
be distinguished by particularly high GW frequencies if the bulk of the GW power comes from a phase when the g-mode frequency is still low.

\begin{figure}
\includegraphics[width=\columnwidth]{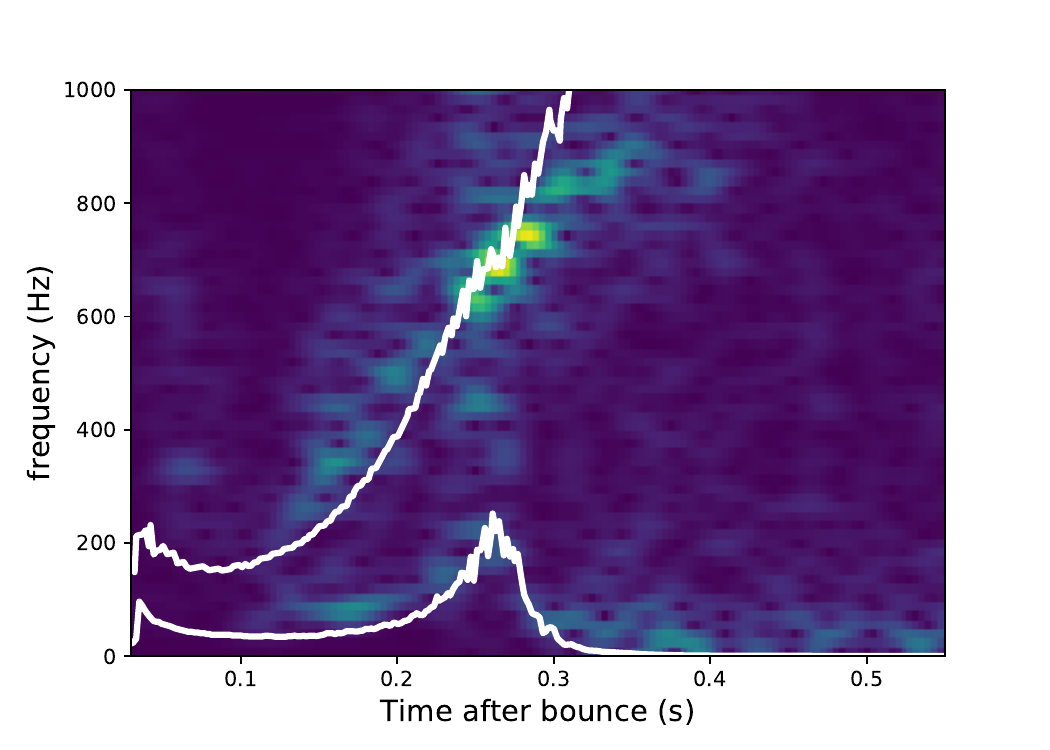}
\includegraphics[width=\columnwidth]{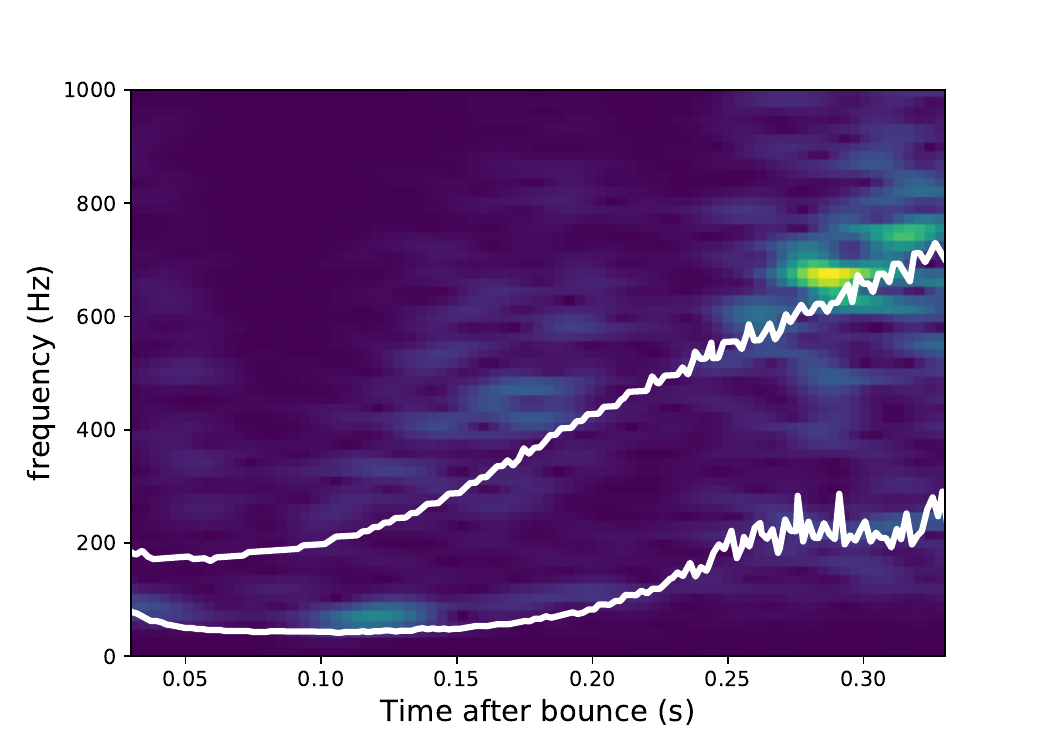}
\caption{Comparison of the spectrogram of the \texttt{z85\_SFHx} model (top), and the \texttt{z100\_SFHx} model (bottom)}(same as in Figure~\ref{fig:spectrograms}) and analytic relations for the g-mode and SASI mode frequencies. The lower white curve shows the GW frequencies predicted by Equation~\ref{eqn:sasi}, accounting for frequency doubling.
The white curve at high frequency shows the GW  frequency predicted by the Universal relations for the $^2g_2$-mode from \citet{2019PhRvL.123e1102T}.
\label{fig:modes}
\end{figure}

The relation between between the mass and radius of the PNS and the mode frequency is largely consistent with semi-analytic estimates
\citep{mueller_13} as in our previous non-rotating models \citep{2019MNRAS.487.1178P, 2020MNRAS.494.4665P} and the universal relations in \citet{2019PhRvL.123e1102T}, especially during the pre-explosion phase. A comparison of the spectrogram of model \texttt{z85\_SFHx} and model \texttt{z100\_SFHx} with the  frequency relation 
for the $^2g_2$ mode from \citet{2019PhRvL.123e1102T} is shown in a separate Figure~\ref{fig:modes} for improved clarity. We show these two models as an example but find the same results for all models.  In fitting the dominant
emission frequency one has to bear in mind that the emitting mode can change character to an f-mode \citep{2018ApJ...861...10M,sotani_20}, but in practice
the suggested scaling of the  $^2g_2$ frequency with
$M/R^2$  also gives a reasonable fit with the f-mode frequency after the character of the mode changes.
During the explosion phase the actual mode frequency from the spectrograms increases more slowly than the analytic scaling relations suggest, which has also been observed previously in 
\citet{mueller_13}. These deviations from the analytic scaling relations in the explosion phase contribute to the dominance of relatively low frequencies in the overall signal in the \texttt{z85} models despite the strong contraction of the PNS on the way to BH formation.

All of our models clearly show low-frequency GW emission at $\lesssim 200\,\mathrm{Hz}$ as well. 
The effects of the different EoS are more significant in the low-frequency GW emission. 
In the case of
model  \texttt{z85\_LS220}, where the shock is revived very early, the low-frequency emission is quite strong, but rather spread out in frequency. It reflects irregular mass motions in the gain region with characteristic time scales of order $\mathcal{O}(10\, \mathrm{ms})$ rather than periodic SASI oscillations. There may also be some confusion between genuine low-frequency emission
from mass motions in the gain region and g-mode emission early on around $100\, \mathrm{ms}$, when the g-mode frequency is still very low.

In the SFHo and SFHx models, the low-frequency GW emission can clearly be attributed to the SASI. The low-frequency emission occurs in a rather clearly defined band in the spectrograms. The
SASI frequency can be approximated as 
\begin{equation}
f_\mathrm{SASI} = \frac{1}{19\,\mathrm{ms}} \left(\frac{R_\mathrm{sh}}{100\,\mathrm{km}}\right)^{-3/2} 
\left[\ln{\left(\frac{R_\mathrm{sh}}{R_\mathrm{PNS}}\right)}\right]^{-1},
\label{eqn:sasi}
\end{equation} 
where $R_\mathrm{sh}$ is the shock radius and $R_\mathrm{PNS}$ is the radius of the PNS \citep{2014ApJ...788...82M}. 
As previously noted by \citet{2017MNRAS.468.2032A}, the SASI emission band in the GW spectograms is located at $2 f_\mathrm{SASI}$ because of frequency doubling similar to GWs from orbiting binaries. Frequency doubling comes about because
after half-cycle of a SASI dipole mode, in which the density distribution
roughly undergoes a spatial reversal $\mathbf{x}\rightarrow -\mathbf{x}$, the mass quadrupole moment $\rho (x_i x_j-\delta_{ij} r^2)$ has already returned to its original value; hence the period of the GW signal is only half the period of the SASI dipole coefficients. Figure~\ref{fig:modes} illustrates (again
for models \texttt{z85\_SFHx} and \texttt{z100\_SFHx}) that in all our models the SASI emission
band is well fit by $2f_\mathrm{SASI}$ up to shock revival.

In the exploding models \texttt{z85\_SFHx} and \texttt{z85\_SFHo}, the SASI emission band increases in frequency with time up to the point of shock revival and again decreases afterwards as the shock expands.
This effect can be seen most clearly in the spectrogram of model \texttt{z85\_SFHx} where shock revival occurs later so that the SASI band can reach a frequency of $\mathord{\sim} 200$\,Hz before the shock is revived and the GW frequency starts decreasing.
As the shock is not revived in the \texttt{z100} models, their low-frequency emission band continues to increase and reaches a frequency of $\sim 400$\,Hz by the end of the simulation.
Therefore, in exploding models, the different EoS result in a clear difference in the low-frequency GW emission,
but since the connection between the microphysical
properties of the EoS and the SASI activity is indirect (through
the shock trajectory)
and may be compounded by progenitor differences, 
it is difficult to directly constrain the EoS based on these
low-frequency signal features.

Overall, the \texttt{z85} models with successful shock revival exhibit larger GW amplitudes in line with previous comparisons of GW emission in exploding and non-exploding models.
The maximum amplitudes (discarding late-time tail signals) for the \texttt{z85} models are $\mathord{\sim} 20\,\mathrm{cm}$, and for the \texttt{z100} models are $\mathord{\sim} 10\,\mathrm{cm}$. The \texttt{z85\_SFHx} and \texttt{z85\_SFHo} models 
develop visible late-time tails, especially model \texttt{z85\_SFHx} with a tail amplitude of over $150\, \mathrm{cm}$.
The tails are due to anisotropic expansion of the shock wave with a positive amplitude indicating a prolate explosion \citep{Murphy_2009}.

%%%%%%%%%%%%%%%%%%%%%%%%%%%%%%%%%%
\subsection{Detection prospects}

\begin{figure}
\centering
\includegraphics[width=\columnwidth]{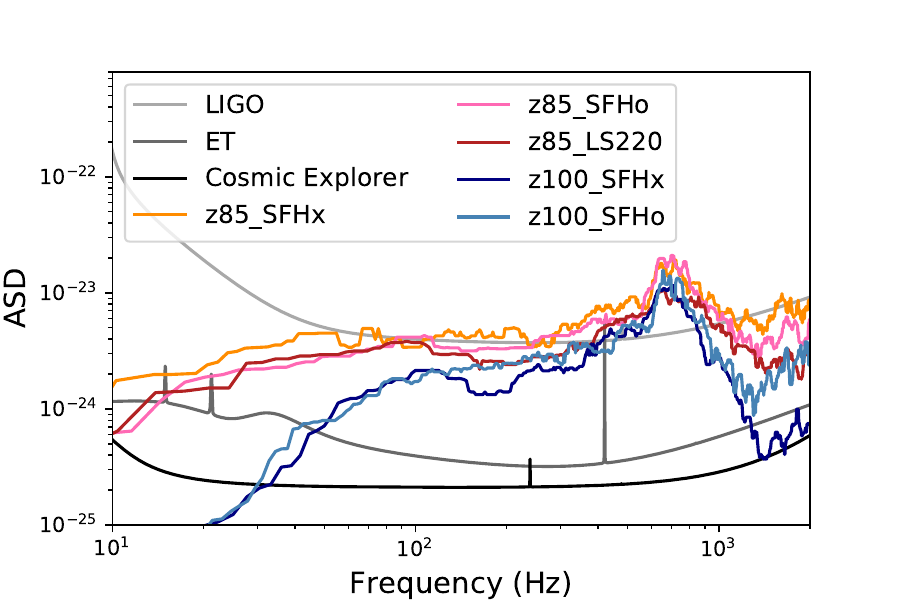}
\caption{The smoothed amplitude spectral density (ASD) 
at $50\, \mathrm{kpc}$ for the five supernova models compared to the
 sensitivity curves of current and future GW detectors. The ASD of the models has been computed assuming an observer in the
 equatorial plane at 
 $(\theta,\varphi)=(90^\circ,90^\circ)$ 
 at a
 distance of $50\, \mathrm{kpc}$. All the models have a maximum amplitude at a frequency of above $\mathord{\sim}700\,\mathrm{Hz}$.
}
\label{fig:asd}
\end{figure}

The amplitude spectral density for all our models at a distance of 50\,kpc is shown in Figure~\ref{fig:asd}. Our models are in a good frequency range for current ground-based GW detectors and future GW detectors with a similar frequency band such as the Einstein Telescope \citep{0264-9381-27-19-194002} and Cosmic Explorer \citep{2017CQGra..34d4001A}. The \texttt{z85} models have stronger low-frequency emission, which will improve their detectability in the Virgo \citep{AdVirgo} and KAGRA \citep{Somiya_2012} GW detectors, which are not as sensitive as LIGO at high frequencies.

We estimate the maximum detectable distance for our models by calculating the matched filter SNR $\rho$,
\begin{equation}
\rho = \frac{\langle s|h \rangle^2} {\langle h|h \rangle\phantom{^2}}\;,
\end{equation}
where $h$ is the waveform and $s$ is the data and the inner product is given by
\begin{equation}
\langle a|b \rangle = 4 \int^{\infty}_{0} \frac{a(f)\,b\!^*\!(f)}{S_{n}(f)}\,\mathrm{d}\!f\;,
\end{equation}
where $S_{n}$ is the power spectral density \citep{1994PhRvD..49.2658C}. 
 As in previous studies, we assume the threshold SNR for detection at the maximum distance is 8 and that the sensitivity of the detectors at the sources sky position is optimal. Below this threshold value it is assumed that the false alarm rate created by detector noise transients will be too large for a confident detection, although it is not currently possible for us to determine the non-Gaussian features of future detectors noise, and knowledge of the sky position and distance of a CCSNe may increase our ability to detect lower SNR signals. In the targeted search for CCSNe during the first and second Observing Runs of LIGO and Virgo, the loudest events had SNRs of $\sim6$ and false alarm rates that indicate they were consistent with background noise \citep{ 2020PhRvD.101h4002A}. Therefore, we assume an SNR of at least 8 will be needed for a signal to be above the background transient noise.  

The results are shown in Table~\ref{tab:detection} for all models and two
different observer directions at $\theta=0^\circ$ (pole) and $(\theta,\varphi)=(90^\circ,90^\circ)$ 
(equator). We show the root sum squared GW amplitude $h_\mathrm{RSS}$ that would be measured by a GW detector for a source at a distance of 10\,kpc. It is defined as
\begin{equation}
h_\mathrm{RSS} = \sqrt{ (h_+ F_+)^2 + (h_\times F_\times)^2 } ,   
\end{equation}
where $h_+$ and $h_\times$ are the two GW polarisations of the signal. $F_+$ and $F_\times$ are the detectors antenna patterns, which are dependent on the source's sky position. We assume them to be equal to 1 which corresponds to an optimal sky position for the source. 
All of the models are detectable at Galactic distances in the Advanced LIGO detector. The \texttt{z85\_SFHx} and \texttt{z85\_SFHo} models have the largest LIGO detection distances with a maximum of $\mathord{\sim} 45\,\mathrm{kpc}$. In a network of advanced detectors, it may be possible to detect these model out to the Large Magellanic Cloud at $48\,\mathrm{kpc}$.  The \texttt{z100\_SFHx} model has the smallest detection distance, which is likely an artefact of the short simulation duration, however. As the model has similar amplitudes to model \texttt{z100\_SFHo}, we expect the distance for the two models would be similar if the model had been simulated for a longer duration. Since the \texttt{z100} models will not collapse to a BH on short time-scales and may yet explode, the detection distances for these models should be considered lower limits.

The models will be detectable at hundreds of kpc in the Cosmic Explorer and Einstein Telescope detectors. The \texttt{z85\_SFHx} model has detectable distances of up to 515\,kpc in Einstein Telescope and 863\,kpc in Cosmic Explorer. This indicates that Cosmic Explorer may detect BH-forming stellar collapse in M31 at $778\,\mathrm{kpc}$. The detection distances of these models may reach up to a few Mpc in a multiple-detector network of third-generation detectors. 

In Table~\ref{tab:detection}, we also show the SNRs in Advanced LIGO 
for each model at 10\,kpc, which range from
17 (\texttt{z100\_SFHx}) to 37 (\texttt{z85\_SFHx}). As we shall see below, such modest SNRs are already enough to spot key features in the time-frequency features of the GW signal.
Splitting the contribution to the SNR from frequencies below and above 350\,Hz, we find that the high-frequency g-modes are the main component of the total SNR. The
direction dependence of the high-frequency and low-frequency contribution to the SNR is modest. 

In the light of a more mature understanding of the time-frequency structure of CCSN GW signals, it is increasingly important to not only address the mere question of detectability or the broad-brush distinction  of different CCSN explosion scenarios
\citep[e.g.,][]{logue_12,powell_16}, but also the problem
of quantitative parameter estimation. Different from the scenario of rotational collapse \citep{2014PhRvD..90d4001A}, quantitative parameter estimation and feature extraction from the post-bounce GW signal is still the subject of active research. Some attempts to extract features from data with realistic noise have already been made by \citet{hayama_15,hayama_18,2019PhRvD..99f3018R}, and recently by
\citet{bizouard_20} based on universal relations for PNS oscillations modes.
As a complementary approach to model-based parameter estimation it is also insightful to directly consider noisy mock data in the time-frequency domain. In order to construct noisy mock spectrograms, we create Gaussian simulated noise for the Advanced LIGO, Einstein Telescope, and Comic Explorer detectors using the ASD curves shown in Figure~\ref{fig:asd}. To create the detector noise, Gaussian points are drawn in the frequency domain around the ASD curves and are then Fourier transformed to create time domain Gaussian noise.  We then add our time domain GW signals to the time domain noise for each detector. Mock spectrograms of signals with LIGO noise at a distance of 5\,kpc for an observer in the equatorial plane
of the spherical polar grid are shown in the right column of Figure~\ref{fig:spectrograms}. 

We find that the characteristic features of the signal, i.e.\ the g-mode and SASI emission band, remain visible 
in the spectrograms even at lower SNRs
than in Figure~\ref{fig:spectrograms} down
to $\mathrm{SNR} \sim 20$. 
Even by eye and with second-generation detectors, the SASI frequency could be pinpointed within $\mathord{\sim}20\%$ for over $100\, \mathrm{ms}$
and the g-mode frequency within $\mathord{\sim}10\%$
at peak emission for a model like \texttt{z85\_SFHx} at a distance of $5\,\mathrm{kpc}$. With third-generation instruments and a higher SNR by a factor of $10\texttt{-}20$, quantitative measurements of mode frequencies will clearly be possible throughout the Milky Way for strong GW emitters.
At lower SNR values, where the features are not visible by eye, it may still be possible to extract the features of the signal using waveform reconstruction techniques \citep{2008CQGra..25k4029K, 2015CQGra..32m5012C}. \citet{2019PhRvD..99f3018R} show they can determine that SASI is present in a spectrogram down to SNR values as low as $\sim 10$. Reconstructing the signal modes in time-frequency space will be essential for relating the properties of the detection to the PNS properties and explosion dynamics.

CCSNe are also expected to produce GWs due to the anisotropic neutrino emission \citep{1997A&A...317..140M, 2009ApJ...697L.133K,2020ApJ...901..108V}. While this signal
component can have amplitudes comparable to the matter
signal, it lies at far lower frequencies where the present
ground-based GW detectors considered here are not very sensitive.

\begin{table*}
\begin{center}
\begin{tabular}{|c|c|c|c|c|c|c|c|c|c|} 
 \hline
Progenitor & EoS & Observer & LIGO  & ET    & CE    & $h_\mathrm{RSS}$ @ 10\,kpc  &  SNR @10\,kpc  & $\mathrm{SNR}_{\mathrm{low}\ f}$ & $\mathrm{SNR}_{\mathrm{high}\ f}$\\
           &     &(position)& (kpc) & (kpc) & (kpc) & ($10^{-23}$)                &                &
                                 &                                  \\ \hline
\texttt{z85}   & SFHx  & pole      & 46  & 479  & 851 & 11.7 & 37  & 18 & 31   \\ \hline
\texttt{z85}   & SFHx  & equator   & 44  & 466  & 825 & 11.6 & 35  & 21 & 28   \\ \hline
\texttt{z85}   & SFHo  & pole      & 44  & 465  & 825 & 11.8 & 36  & 18 & 29   \\ \hline
\texttt{z85}   & SFHo  & equator   & 42  & 427  & 783 & 11.1 & 34  & 18 & 28   \\ \hline
\texttt{z85}   & LS220 & pole      & 37  & 386  & 690 & 8.51 & 30  & 11 & 27   \\ \hline
\texttt{z85}   & LS220 & equator   & 30  & 309  & 556 & 7.89 & 24  & 14 & 18   \\ \hline
\texttt{z100}  & SFHx  & pole      & 21  & 216  & 381 & 5.08 & 17  & 8  & 15   \\ \hline
\texttt{z100}  & SFHx  & equator   & 21  & 235  & 408 & 5.34 & 17  & 8  & 16   \\ \hline
\texttt{z100}  & SFHo  & pole      & 24  & 253  & 439 & 6.81 & 19  & 8  & 17   \\ \hline
\texttt{z100}  & SFHo  & equator   & 26  & 282  & 492 & 7.05 & 22  & 11 & 19   \\ \hline
\end{tabular}
\caption{Summary of GW emission for all models for observers at the pole ($\theta=0^\circ$)
and in the equatorial plane at
$(\theta,\varphi)=(90^\circ,90^\circ)$. Columns LIGO, ET and CE show the maximum distances in kpc based on a threshold  matched filtering signal-to-noise ratio (SNR) of 8 in the LIGO, Einstein Telescope, and Cosmic Explorer detectors, respectively. $h_\mathrm{RSS}$ is the root-sum-squared amplitude at 10\,kpc.
Column SNR shows the total SNR of the signal in LIGO for a source at 10\,kpc. 
The last two columns 
$\mathrm{SNR}_{\mathrm{low}\ f}$ and $\mathrm{SNR}_{\mathrm{high}\ f}$
show the SNR calculated using only the signal at frequencies below and above 350\,Hz. }
\label{tab:detection}
\end{center}
\end{table*}

%%%%%%%%%%%%%%%%%%%%%%%%%%%%%%%%%%%%%%%%%%%%%%%%%%
%%%%%%%%%%%%%%%%%%%%%%%%%%%%%%%%%%%%%%%%%%%%%%%%%%
\section{Conclusions}
\label{sec:conclusion}

In recent years, a greater understanding of CCSN explosions has been reached through self-consistent 3D simulations.
One of the challenges for 3D CCSN models is now to scan the progenitor parameter space and broadly survey the outcomes of stellar collapse in terms of remnant and explosion properties.
More models are also needed to build a more extensive bank of gravitational waveform predictions in order to aid and inform future detections of CCSNe in GWs.
In the light of recent GW detections, the
final collapse of progenitors in the pulsational pair instability regime and 
origin of the most massive BHs produced by CCSNe are of particular interest and need
to be explored more thoroughly by first-principle supernova models.

For this reason, we performed CCSN simulations of two very massive progenitors  with the neutrino hydrodynamics code \textsc{CoCoNuT-FMT}. The progenitor stars we use are $85\,\mathrm{M}_\odot$ and $100\,\mathrm{M}_\odot$ Pop-III stars from the lower end of the pulsational pair instability regime. 
We used three different nuclear EoS (LS220, SFHx, and SFHo EoS) to examine the EoS sensitivity
of the dynamics and GW emission of supernovae
from very massive progenitors. 

In all of the $85\, \mathrm{M}_\odot$ models, the
shock is revived at relatively early post-bounce times.  Because of their very massive cores, these models then form BHs
within a few hundreds of milliseconds
after shock revival, however. These
findings provide further  indication for
precipitous shock revival in progenitors with high compactness
\citep{2018ApJ...865...81O,burrows_20}, which could then develop into fallback supernovae \citep{2018ApJ...852L..19C} with partial envelope ejection or completely collapse to a BH. The model with the LS220 EoS explodes the earliest at 0.17\,s after core-bounce and quickly forms a BH at 0.29\,s after bounce. The model with the SFHx EoS, which supports the highest maximum mass, takes the longest time until shock revival, but is also the last to collapse a BH and reaches the largest diagnostic energy of $2.7\times10^{51}\,\mathrm{erg}$ due to the longer accretion time. Even in this case, the energy is not sufficient to shed the entire envelope and BH collapse by fallback is unavoidable. Longer simulations are required to decide whether the incipient explosions lead to partial mass ejection or are eventually stifled. For the most energetic explosion with the SFHx EoS, we estimate a final BH mass in the range of $30.7\texttt{-}34.4\,\mathrm{M}_\odot$.

The $100\, \mathrm{M}_\odot$ models did not explode before the end of the simulation, but heating conditions are already close to runaway shock expansion so
that these models would likely explode before BH collapse.  Further simulations will be needed to determine which stars in the pair instability regime will quietly form BHs during their final collapse, and which ones will undergo early or late shock revival before BH formation and perhaps shed part of the envelope.

We determined the GW emission for all of our models. The GW spectrograms exhibit familiar features with a high-frequency g-mode emission band, and all of the models also have quite strong low-frequency emission. In the SFHx and SFHo models, the low-frequency emission clearly stems from  strong SASI emission. In the $85\, \mathrm{M}_\odot$ models, the frequency of the SASI emission band increases with time up to the point of shock revival where the SASI disappears. In the $100\, \mathrm{M}_\odot$ models, the SASI emission band continuously increases in frequency and remains present throughout the simulations even though GW amplitudes decline after $400\, \mathrm{ms}$.  The time-integrated GW spectrum peaks at frequencies 
of $650\texttt{-}710\, \mathrm{Hz}$, 
within the sensitivity range of current and third generation GW detectors, which is somewhat higher than the detectors peak sensitivity range, 
but not unusually high compared to CCSN models of less massive progenitors.

Overall, the GW emission from these very massive progenitors is strong and favourable for detection. We obtain maximum detection 
distances of up to $46 \, \mathrm{kpc}$ with Advanced LIGO. Bearing in mind that some of the waveforms are still incomplete, the GW signals
from these pulsational pair instability models 
should be detectable throughout the Galaxy and perhaps in the  Large Magellanic Cloud with present-day GW detector networks. The $85\, \mathrm{M}_\odot$ models with the SFHo and SFHx EoS would be detectable out to M31 in Cosmic Explorer.  We demonstrated that the g-mode and SASI emission bands can be identified in noisy spectrograms even by eye for moderately high
SNRs of $20$, which are easily reached for events in the Milky Way and its satellites in third-generation instruments. This underscores the potential for measuring the dynamics of quiet BH collapse or weak explosions with GWs in the next decades.

To date there are some potential candidates for pulsational pair instability supernovae but no confirmed observations \citep{2017Natur.551..210A,2018ApJ...863..105W, 2019ApJ...881...87G}. Some theoretical studies have made predictions of a lower limit on the rate of pulsational pair instability supernovae of $\sim 0.1 \mathrm{Gpc}^{-3} \mathrm{yr}^{-1}$ at redshift zero \citep{2019ApJ...882..121S}. Pulsational pair instability supernovae are unlikely to occur in our Galaxy, as they are only expected to occur in low metallicity environments, however it is possible they may be detected in the local group by the next generation of GW detectors.

%%%%%%%%%%%%%%%%%%%%%%%%%%%%%%%%%%%%%%%%%%%%%%%%%%
%%%%%%%%%%%%%%%%%%%%%%%%%%%%%%%%%%%%%%%%%%%%%%%%%%
\section*{Acknowledgements}

We thank KaHo Tse for providing the code for the periodogram used to analyse the pre-supernova evolution of model \texttt{z85}.
We thank Kei Kotake for helpful comments. 
The authors are supported by the Australian Research Council (ARC) Centre of Excellence (CoE) for Gravitational Wave Discovery (OzGrav) project number CE170100004.  JP is supported by the ARC Discovery Early Career Researcher Award (DECRA) project number DE210101050.  BM is supported by ARC Future Fellowship FT160100035.  AH is supported by the ARC CoE for All Sky Astrophysics in 3 Dimensions (ASTRO 3D) project number CE170100013.
We acknowledge computer time allocations from Astronomy Australia Limited's ASTAC scheme and the National Computational Merit Allocation Scheme (NCMAS).
Some of this work was performed on the Gadi supercomputer with the assistance of resources and services from the National Computational Infrastructure (NCI), which is supported by the Australian Government.  Some of this work was performed on the OzSTAR national facility at Swinburne University of Technology.  OzSTAR is funded by Swinburne University of Technology and the National Collaborative Research Infrastructure Strategy (NCRIS).

%%%%%%%%%%%%%%%%%%%%%%%%%%%%%%%%%%%%%%%%%%%%%%%%%%
\section*{Data Availability}

The data from our simulations will be made available upon reasonable requests made to the authors. 


\bibliographystyle{mnras}
\bibliography{ms} 

%%%%%%%%%%%%%%%%%%%%%%%%%%%%%%%%%%%%%%%%%%%%%%%%%%

% Don't change these lines
\bsp	% typesetting comment
\label{lastpage}
\end{document}